\documentclass[a4paper,aps,prb,reprint,showpacs,superscriptaddress]{revtex4-1}
\usepackage{amsmath,amssymb,mathrsfs}
\usepackage[pdftex]{graphicx}
\usepackage{epstopdf}
\usepackage{xspace}
\usepackage{color}

\usepackage[latin9]{inputenc}
\usepackage{enumitem} % for A,B enumeration
\usepackage{esint}
\usepackage{xargs}[2008/03/08]
\usepackage[caption=false,size=Large]{subfig} % subig, NOT:subcaption package incompatible with RevTeX !!!
\usepackage{etoolbox} % make user-defined commands robust: (re)newrobustcmd
\robustify{\sum}
\robustify{\int}
\robustify{\nonumber}
\robustify{\cite}
\robustify{\footnote}

\newrobustcmd{\ket}[1]{\left|#1\right\rangle}
\newrobustcmd{\bra}[1]{\left\langle#1\right|}
\newrobustcmd{\braket}[1]{\left\langle#1\right\rangle}
\newrobustcmd{\sgn}[1]{\text{sgn}(#1)\,}
\newrobustcmd{\ii}{\text{i}}
\newrobustcmd{\dd}{\text{d}}
\newrobustcmd{\abs}[1]{\left|#1\right|}
\newrobustcmd{\comm}[2]{\left[#1,#2\right]}
\newrobustcmd{\anticomm}[2]{\left\{#1,#2\right\}}
\newrobustcmd{\average}[1]{\langle#1\rangle}

\newrobustcmd{\Cite}[1]{Ref.~\onlinecite{#1}}
\newrobustcmd{\Cites}[1]{Refs.~\onlinecite{#1}}
\newrobustcmd{\Tab}[1]{Table~\ref{#1}}
\newrobustcmd{\Fig}[1]{Fig.~\ref{#1}}  
\newrobustcmd{\Figs}[1]{Figs.~\ref{#1}}  
\newrobustcmd{\Eq}[1]{Eq.~(\ref{#1})}
\newrobustcmd{\Eqb}[1]{Equation~(\ref{#1})} 
\newrobustcmd{\Eqs}[1]{Eqs.~(\ref{#1})}  
\newrobustcmd{\eq}[1]{(\ref{#1})}  
\newrobustcmd{\App}[1]{App.~\ref{#1}}
\newrobustcmd{\Sec}[1]{Sec.~\ref{#1}}

\DeclareMathOperator{\tr}{Tr}

\newrobustcmd{\icot}{inelastic tunneling\xspace}
\newrobustcmd{\set}{resonant tunneling\xspace}
\newrobustcmd{\coset}{COSET\xspace}

\usepackage{simplewick}
\newcommandx\wick[5][usedefault, addprefix=\global, 1=1ex]{\acontraction[#1]{#2}{#3}{#4}{#5}}

\newcommand{\new}[1]{#1}                    % UNCOLORED by default (so senior Utrecht professors don't mess it up)
\newcommand{\New}[1]{#1}                    % UNCOLORED by default (so senior 

\newcommand{\col}[4]{{\begin{bmatrix}#1 \\ #2 \\ #3 \\ #4 \end{bmatrix}}}
\begin{document}
%%%%%%%%%%%%%%%%%%%%%%%%%%%
\title{Charge fluctuations in nonlinear heat transport}
%%%%%%%%%%%%%%%%%%%%%%%%%%%
\author{Niklas M. Gergs}
\affiliation{Institute for Theoretical Physics, Center for Extreme Matter and Emergent Phenomena, Utrecht University, Leuvenlaan 4, 3584 CE Utrecht, The Netherlands}
\author{Christoph B. M. H\"orig}
\affiliation{Institute for Theoretical Physics, Center for Extreme Matter and Emergent Phenomena, Utrecht University, Leuvenlaan 4, 3584 CE Utrecht, The Netherlands}
\author{Maarten R. Wegewijs}
\affiliation{Peter Gr\"unberg Institut, Forschungszentrum J\"ulich, 52425 J\"ulich, Germany}
\affiliation{Institute for Theory of Statistical Physics, RWTH Aachen University, 52056 Aachen, Germany}
\affiliation{JARA--Fundamentals of Future Information Technology}
\author{Dirk Schuricht}
\affiliation{Institute for Theoretical Physics, Center for Extreme Matter and Emergent Phenomena, Utrecht University, Leuvenlaan 4, 3584 CE Utrecht, The Netherlands}
\date{26 May 2015}
\pagestyle{plain}

\begin{abstract}
We show that charge fluctuation processes are crucial for the nonlinear heat conductance through an interacting nanostructure, even far from a resonance. We illustrate this for an Anderson quantum dot accounting for the first two leading orders of the tunneling in a \New{master equation.} The often made assumption that off-resonant transport proceeds entirely by virtual occupation of charge states, underlying exchange-scattering models, can fail dramatically for heat transport. The identified energy-transport resonances in the Coulomb blockade regime provide new qualitative information about relaxation processes, for instance by strong \emph{negative} differential heat conductance \New{relative to the} heat current.
These can go unnoticed in the charge current, making nonlinear \emph{heat-transport spectroscopy} with energy-level control a promising experimental tool.
\end{abstract}
%05.60.Gg Transport processes: Quantum transport
%73.23.Hk Electronic transport in mesoscopic systems: Coulomb blockade; single-electron tunneling
%73.63.Kv Electronic transport in nanoscale materials and structures: Quantum dots
%73.63.-b Electronic transport in nanoscale materials and structures
%73.50.Lw Electronic transport phenomena in thin films: Thermoelectric effects
\maketitle

Recently the experimental investigation of heat transport on the nanoscale has become possible~\cite{Reddy07,Baheti08,Widawsky12}. 
\new{These measurements are accurate enough to investigate the heat dissipation in molecular junctions with conductances as low as~\cite{Lee-13} $10^{-3}e^2/h$.}
%These measurements are accurate enough to determine the asymmetry of heat dissipation and its relation to the electronic transmission characteristics for molecular junctions with conductances as low as~\cite{Lee-13} $10^{-3}e^2/h$. 
%Although experimentally even more challenging, additionally 
\new{Additionally integrating energy-level control into thermoelectric junctions, e.g., by mechanical~\cite{Temirov08} gating, does not seem out of reach and just recently, electrically-gated thermoelectric nanojunctions have been demonstrated~\cite{Kim-14}.}
\New{Here, by analysing the generic effects of Coulomb interactions on the nonlinear heat transport in nanoscale systems, we will show that this is very promising.}

\begin{figure}[t]  
\centering\includegraphics[width=1.0\columnwidth]{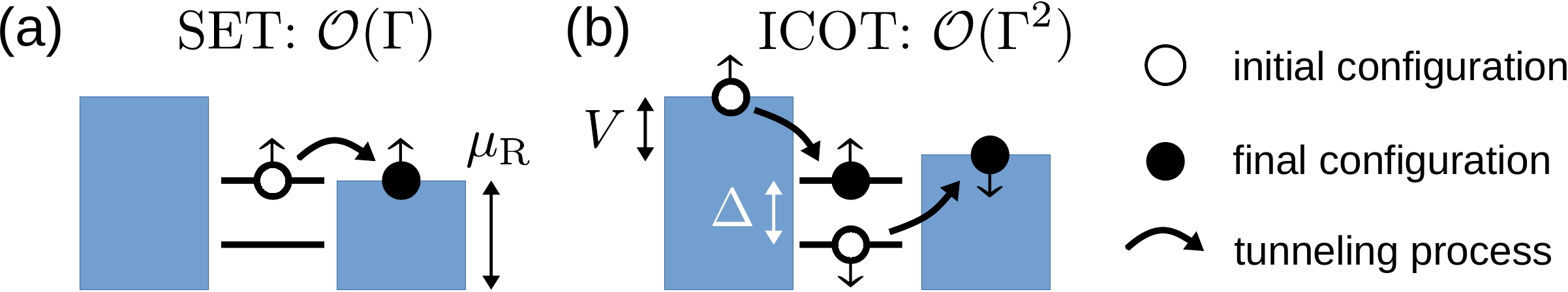}
\caption{\new{(Color online)
Examples of tunneling processes between electrodes (blue) and discrete quantum levels (black) (a) of first and (b) of second order in the tunneling rate $\Gamma$.
%(a) SET $\mathcal{O}(\Gamma)$.
%(b) ICOT $\mathcal{O}(\Gamma^2)$.
}
\label{fig:0}
}
\end{figure}

\begin{figure}[t]  
\centering\includegraphics[width=\columnwidth]{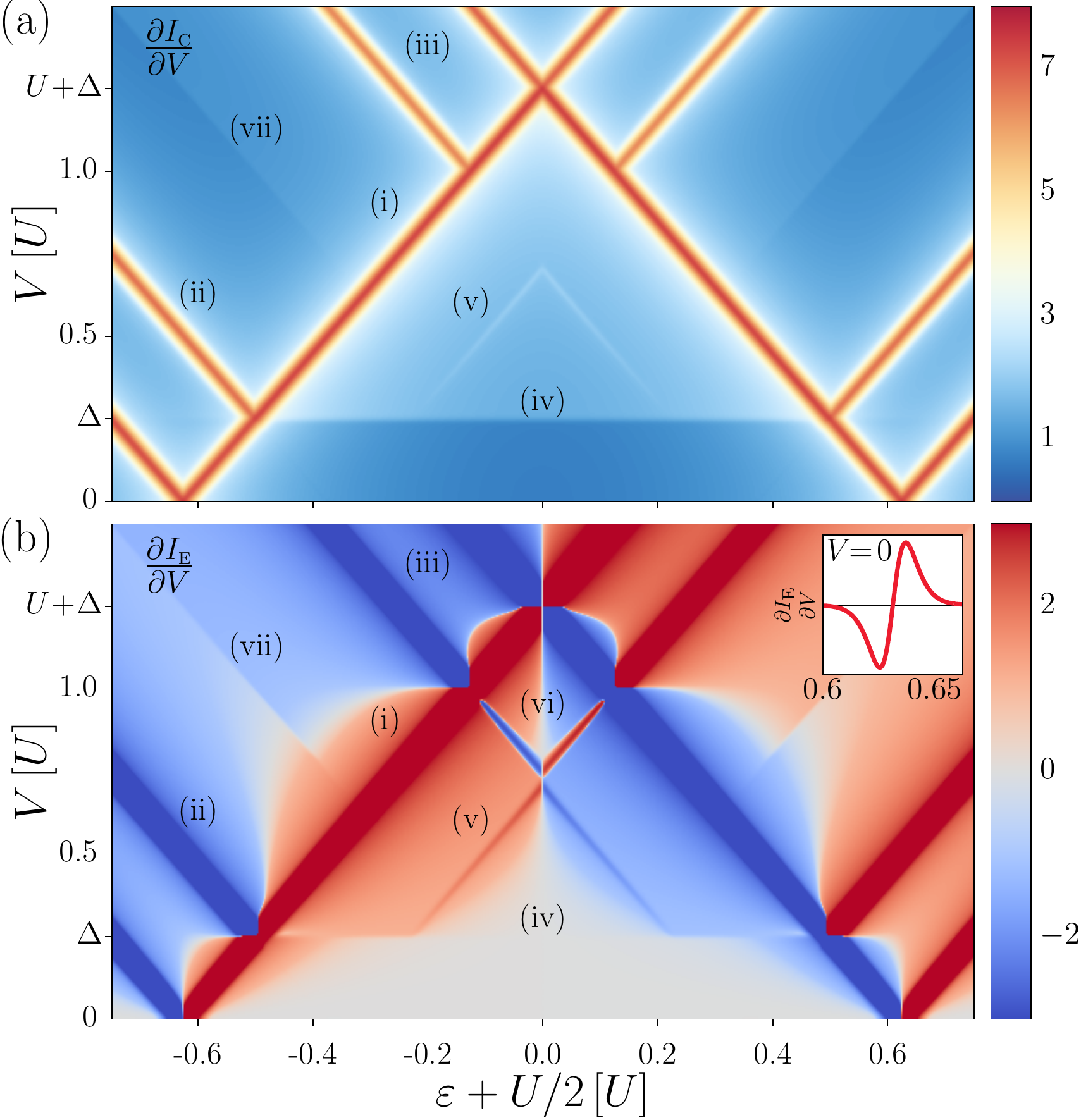}
\caption{(Color online) Transport through a quantum dot with
\New{interaction} $U=\tfrac{1}{3} \,10^3\,T$, \New{inelastic excitation} $\Delta=\tfrac{1}{4}U \approx 83.3\,T$,
\New{and tunnel coupling} $\Gamma=\tfrac{1}{3} \,10^{-2}\,T$.
(a) Charge conductance  $\log_{10} \left( [\partial I_\text{C}/\partial V] / [\Gamma^2/U^2] \right)$ and (b) energy conductance $\text{slog}_{10}\left( [\partial (I_\text{E}/\partial V)] / [\Gamma^2/U] \right)$
\New{using} the signed log,
 $\text{slog}_{10}(x) := \sgn{x}\log_{10}(a|x|)$ for $a|x| \geq 10$ \New{with $a=20$, linearized to} $\text{slog}_{10}(x):= ax/10$ for $a|x| \leq 10$.
Labels (i)--(vii) indicate the features discussed in the text but are not labeled at horizontally mirrored positions.
Inset to (b): Linear energy conductance [a.u.] versus $\varepsilon+U/2$ around the right SET resonance.
%In (b) we plot the signed log, $\text{slog}_{10}(x) := \sgn{x}\log_{10}(a|x|)$ for $a|x|>10$, which is linearized near zero, $\text{slog}_{10}(x):= ax/10$ for $a|x|<10$ using $a=20$.
\label{fig:1}}
\end{figure}

Interaction effects have \new{long} been probed using gate controlled \emph{charge}-current spectroscopy, a well-developed experimental tool to access the discrete quantum levels of nanostructures. Two prominent features in the charge current driven by a source-drain voltage underpin this successful method. The first is resonant or single-electron tunneling (SET) which  depends on the level position relative to the electro-chemical potential, \new{$\mu_\text{R}$ in~\Fig{fig:0}(a):
An electron jumps into or out of an orbital level directly
leading to a real change of its occupancy.}
The current shows sharp steps as new resonant transport processes are switched on with increasing bias.
These processes are \new{routinely} identified in a three-terminal setup by plotting the charge conductance as function of the applied bias $V$ and the gate voltage,
as exemplified in \Fig{fig:1}(a).
Two-terminal measurements, e.g., using a scanning probe, correspond to line traces through such a plot. The second type of resonance is independent of the level position and appears as a horizontal line at $V=\Delta$ since it originates in the inelastic excitation by an energy $\Delta$ at fixed local electron number on the nanostructure. This off-resonant feature requires a second-order tunneling process in which an electron ``scatters through'', other charge states being only visited virtually, \new{see \Fig{fig:0}(b).
This is} known as inelastic electron tunneling (IETS)~\cite{Lambe68,Heinrich04} or inelastic cotunneling (ICOT)~\cite{GlazmanMatveev90,Averin90,DeFranceschi01,Paaske06}.
%Under appropriate conditions this \icot resonance can be described in terms of effective exchange- and potential-scattering amplitudes.
\new{This \icot resonance}  develops into a nonequilibrium Kondo resonance for low $\Delta$ and low temperatures~\cite{Rosch03,Kehrein05,Schoeller09b}
\new{which is much sharper}~\cite{DeFranceschi01,Paaske06} than the \set feature corresponding to $\Delta$, providing \new{better} access to a range of physical phenomena \emph{in situ}: an electronic level splitting (e.g., in a semiconductor nanostructure~\cite{Zumbuehl04}, carbon nanotube~\cite{Sapmaz05b,Huettel09}, or a dopand atom~\cite{Zwanenburg13}), a quantized vibrational frequency~\cite{Osorio07a}, or a spin-splitting due to a magnetic field~\cite{Heinrich04}, exchange interaction~\cite{DeFranceschi01,Paaske06,Grose08}, magnetic anisotropy (e.g., in molecules~\cite{Parks07,Zyazin10} or ad-atoms~\cite{Loth10}), or spin-orbit coupling~\cite{Jespersen11}.

Thermoelectric transport has also been investigated within the two above mentioned physical transport pictures. Theory mostly focused on the thermopower in the linear-response regime. This includes the study of \set~\cite{BeenakkerStaring92}, \icot~\cite{TurekMatveev02,Koch04a,Kubala06} and Kondo processes~\cite{CostiHewson93,KimHershfield02,Scheibner05,CostiZlatic10}. Works addressing the nonlinear regime have either applied effective single-particle \new{descriptions~\cite{Galperin07,LopezSanchez13,Whitney13,Svensson13,Zotti14} or} focused on thermoelectric devices close to resonance assuming weak tunneling~\cite{Esposito09,Leijnse10,Wang12} or weak \new{Coulomb} interaction~\cite{Kennes13b}. The heat current has received much less attention~\cite{Esposito09,Leijnse10,Wang12,Kennes13b}. A classification of \emph{nonlinear heat-transport} features for a strongly interacting nanostructure going beyond weak tunneling, matching that of charge transport~\cite{Schleser05,Koch06,Leijnse08a}, still seems to be missing. 
\new{This is important both for scanning probe setups~\cite{Reddy07,Baheti08,Widawsky12,Lee-13,Temirov08} as well as thermoelectric setups~\cite{Scheibner05,Kim-14} with energy-level control.} 
%This is important not only for thermoelectric setups~\cite{Scheibner05,Kim-14} with energy-level control, but also for scanning probe setups~\cite{Reddy07,Baheti08,Widawsky12,Lee-13} in which the level position plays a crucial \new{role~\cite{Temirov08}.}
%and may sometimes be gated mechanically~\cite{Temirov08}. 
In this Rapid Communication we address this problem and show that the heat current driven by a nonlinear electric and/or thermal bias contains new qualitative information and deviates in a striking way from the charge transport, both in sign and amplitude. Its dependence on the level position reveals that relaxation processes of first order \new{[\Fig{fig:0}(a)]} in the tunneling can be very important for heat transport \emph{far from resonance} (i.e., energy detuning larger than temperature).
%not within the tunnel or temperature broadening). 
The \New{crucial} competition with an inelastic second-order excitation at finite voltage bias $V=\Delta$ \new{[\Fig{fig:0}(b)]} leads to real occupation of more than one charge state
\New{and is missed by inelastic transport theories relying on effective exchange- and potential-scattering amplitudes.}

\emph{Model and method.}---To illustrate the generic picture of \New{nonlinear} thermoelectric transport through an interacting nanoscale object we \New{analyze a} resonant level with strong Coulomb interaction and a well-defined spin-flip excitation $\Delta$ due to an external field. \New{It is described by the Anderson quantum dot model $H_\text{tot}=H_\text{d}+H_\text{res}+H_\text{tun}$ which also} suffices to classify nonlinear thermoelectric transport features %that we obtained 
for more complex models~\cite{supplement}. These features should generally be observable in a range of nanostructures, at least for large level spacings and quasi-regular electron filling of energy shells. The dot is described by $H_{\text{d}}=\sum_{\sigma}(\varepsilon+\sigma \Delta/2)d_{\sigma}^{\dagger}d_{\sigma}+U N(N-1)/2$, where $d_\sigma$ with $\sigma=\uparrow,\, \downarrow$ are the electron operators on the dot. Here $\varepsilon=(\varepsilon_\uparrow + \varepsilon_\downarrow)/2$ is the orbital energy level and $\Delta=\varepsilon_\uparrow - \varepsilon_\downarrow$  denotes the energy of a local spin excitation for fixed $N=1$ due to a magnetic field, where $N=\sum_\sigma d^\dag_\sigma d_\sigma$ is the electron number. Furthermore, $U$ is the strong Coulomb energy penalty paid when counting $N=2$ electrons in the shell. The electrodes, indexed by $\alpha$=L,R, are described as noninteracting reservoirs, $H_\text{res}=\sum_\alpha H_\text{res}^\alpha=\sum_{\alpha k\sigma}\epsilon_k\,c_{\alpha k\sigma}^\dagger c_{\alpha k\sigma}$, with electron operators $c_{\alpha k\sigma}$. We allow for \New{a} nonlinear voltage bias $V$ between the reservoirs through their electrochemical potentials $\mu_\text{\text{L},\text{R}}=\pm V/2$
\New{with temperatures $T_\text{L}=T_\text{R}=T$. (We comment on nonlinear thermal bias effects~\cite{supplement} later on.)}
%as well as a nonlinear thermal bias through their temperatures $T_\text{\text{L},\text{R}}$.
The tunnel coupling has the generic form $H_\text{tun}= t \sum_{k\alpha\sigma} (c_{k\alpha\sigma}^{\dagger}d_{\sigma}+\text{h.c.})$, the bare resonance width is given by the tunnel rate $\Gamma=2\pi\nu_0 t^2$, with $\nu_0$ the density of states in the reservoirs and we set $e=\hbar=k_\text{B}=g\mu_\text{B}=1$.

\begin{figure*}[t]  
\centering\includegraphics[width=2\columnwidth]{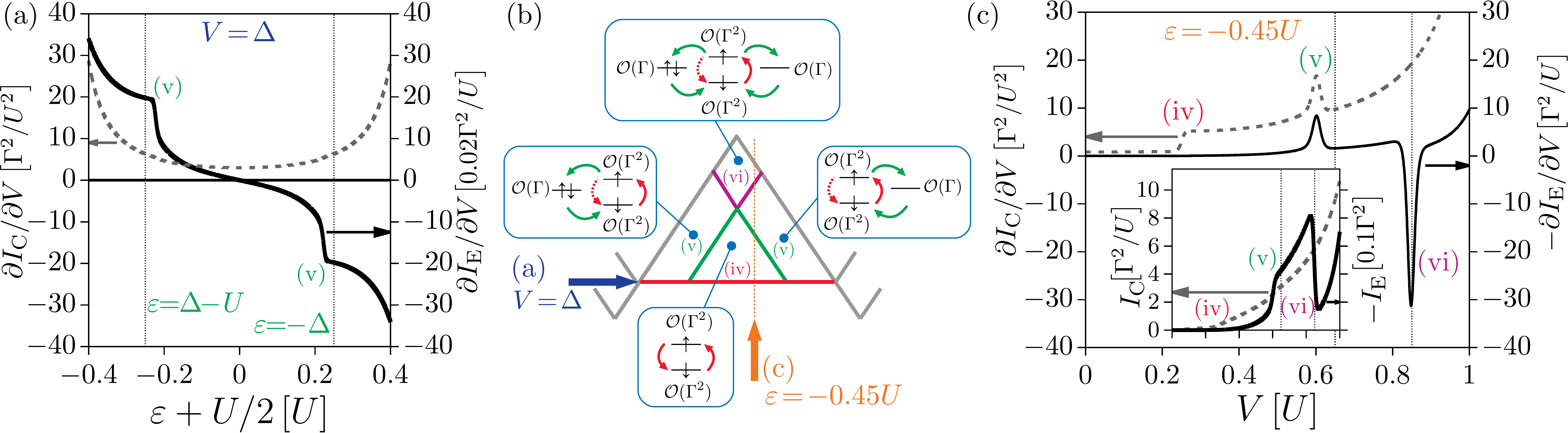}
\caption{(Color online) \New{Explanation of \Fig{fig:1}.
In (b) we sketch  the stability diagrams of \Fig{fig:1} showing by blue and orange arrows where the cuts  in (a) and (c) are taken. (a) and (c) show} $\partial I_\text{C}/\partial V$ (dashed) and $\partial I_\text{E}/\partial V$ (solid) as function of the level position $\varepsilon$ for fixed $V$ \emph{and vice-versa}, respectively, \New{and} the vertical dotted lines indicate COSET resonance positions.
\New{In (c) we plot the negative of $\partial I_\text{E}/\partial V$ and  $I_\text{E}$ for clarity.
In (b), the} boundaries of various regimes (iv)-(vi) and the corresponding  processes \New{discussed} are shown the inset boxes. The gray lines are the well-known SET resonances. The horizontal red line is the $V=\Delta$ threshold for ICOT excitation \New{[\Fig{fig:0}(b)]} shown in the inset box to (iv). When crossing from \New{(iv)} either of the green lines, \New{a \emph{single}} two-step relaxation path is switched on (COSET), colored green in the inset to regimes (v). When subsequently crossing the purple lines (vi) \emph{both} these green relaxation paths become active as shown in the inset box to (vi).
%The energy current is negative since hole fluctuations 
 \label{fig:2}}
\end{figure*}
We will only consider the stationary currents entering the right \New{reservoir.}
The charge current is $I_\text{C}=\left\langle \frac{\dd}{\dd t}N^\text{R}_\text{res}\right\rangle$, where $N^\text{R}_\text{res}$ is the electron number operator of the reservoir $\alpha=\text{R}$. Similarly, the energy current is defined via $I_\text{E}=\left\langle \frac{\dd}{\dd t}H_\text{res}^\text{R}\right\rangle$. The measurable heat current can be obtained via~\cite{Giazotto06} $I_\text{Q}=I_\text{E}-\mu_\text{R} I_\text{C}$. Since experimentally the way the voltage is applied is known (here $\mu_{\text{R}} =-V/2 =- \mu_{\text{L}}$) and the conserved charge current is available, the conversion from $I_\text{Q}$ to $I_\text{E}$ amounts to a simple background subtraction. Below we focus on the contribution $I_\text{E}$ since it contains all interesting physical features. Also, $I_\text{C}$ and $I_\text{E}$ are more easily compared, highlighting the differences between charge and heat transport most directly, in particular the bias and gate voltage dependence on which we focus here. We use $U$ as unit of energy; in experiments it is readily obtained from the height of the Coulomb diamond (cf. \Fig{fig:1}) and ranges from 0.1-10 meV in semiconductor~\cite{Zumbuehl04} and carbon nanotube quantum dots~\cite{Sapmaz05b,Huettel09} to 10-100 meV in molecular~\cite{Osorio07a} and atomic quantum dots~\cite{Zwanenburg13,Loth10}. The effects in the currents that we focus on below scale as $\partial I_\text{E}/\partial V  \propto U\,\partial I_\text{C}/\partial V = \Gamma^2/U$ for the parameter regime of interest $\Gamma\ll U$ when fixing $\Delta$ relative to $U$. Estimations based on this~\cite{supplement} indicate that the predicted energy currents may be in range of experimental resolution of tens of nW, in particular in molecular junctions.

The currents and the underlying nonequilibrium dot-state occupations are calculated using a reduced density-operator transport theory~\cite{Schoeller09a,Saptsov12a} accounting for the strong local interaction $U$. This approach is perturbative in the tunneling rates and well-controlled in the regime $\Gamma\ll T$.
% while there is no restriction on the other parameters
\new{While} keeping this restriction we recover~\cite{supplement} for $U=0$ the corresponding results of the Landauer approach\cite{Lee-13, Zotti14}. We go beyond standard approaches by including the competition of all \new{tunneling rates $\mathcal{O}(\Gamma)$ and $\mathcal{O}(\Gamma^2)$ [\Fig{fig:0}]} into the \New{stationary master equation $\dot{p}=0=Wp$ for the occupations $p=(p_0,p_\uparrow,p_\downarrow,p_{\uparrow\downarrow})$, see \Cite{supplement} for more details of the calculations~\cite{Leijnse08a,Leijnse09a,Koller10} of the transition rate matrix $W$ and the current.} We focus on the dominant energy dependence introduced by the interacting quantum dot, \New{assuming a flat spectral density in the wide-band limit for the electrodes.}

\emph{$\mathcal{O}(\Gamma)$ effects.}---\New{Already a first glance at the charge and energy conductance plotted in \Fig{fig:1}(a) and (b), respectively, reveals that the energy transport spectrum is much richer:
there is a significant gain in the contrast due to its many sign changes.}
% Even at zero thermal bias, the energy transport spectrum is much richer than that for the charge: Already a first glance at the charge and energy conductance plotted in \Fig{fig:1}(a) and (b), respectively, reveals a significant gain in the contrast in the latter due to its many sign changes.
In these plots the stage is set by \set features due to processes of $\mathcal{O}(\Gamma)$ \new{[\Fig{fig:0}(a)]} which are well understood~\cite{supplement}. These occur when one of the four single-electron addition energies $\varepsilon_\sigma$, and $\varepsilon_\sigma+U$ ($\sigma=\uparrow,\, \downarrow$) matches $\mu_{\text{L},\text{R}}=\pm V/2$. This happens, e.g., at the lines labeled (i)--(iii) in \Fig{fig:1}. Indicated by (i) are \set transitions between the ground states of subsequent electron numbers $N$ of the dot ($0\to 1$ and $1\to 2$). As shown in the inset in \Fig{fig:1}(b), there are sawtooth-shaped resonances~\cite{BeenakkerStaring92} in  $\partial I_\text{E}/\partial V$ as function of the level position $\varepsilon$ corresponding to the Coulomb peaks in $\partial I_\text{C}/\partial V$ associated with one-electron processes $\ket{\downarrow}\to \ket{0}$ and $\ket{\downarrow} \to \ket{\uparrow\downarrow}$, respectively.
The sign change in $\partial I_\text{E}/\partial V$ \new{reflects that} excess energy is carried by electrons or holes. This basic energy transport feature reappears at several positions in \Fig{fig:1}(b), e.g., also when at (ii) a \set process additionally excites the dot or at (iii) such a process starts off in the $N=1$ excited state $\ket{\uparrow}$,
\New{the process of \Fig{fig:0}(a).} 
\new{The broadening of these \set lines is determined by the temperature for $T\gg \Gamma$.}

\emph{$\mathcal{O}(\Gamma^2)$ effects.}---Qualitative differences show up inside the \New{central} off-resonant regime---opened up by the Coulomb interaction $U$---where the simple resonant picture \New{just} discussed breaks down. Here coherent electron-hole processes of $\mathcal{O}(\Gamma^2)$ \New{that leave $N$ fixed, such as \Fig{fig:0}(b),} become important as well. These give rise to \emph{qualitatively} new effects not captured by $\mathcal{O}(\Gamma)$ master equations or even approaches that also include tunnel broadening and shifts~\cite{Svensson13}. For voltages $V\lesssim \Delta =\varepsilon_\uparrow - \varepsilon_\downarrow$ only elastic $\mathcal{O}(\Gamma^2)$ tunneling processes are possible which produce a smooth nonexponential background in both $\partial I_\text{C}/\partial V$ (qualitatively similar to that found for metallic islands~\cite{TurekMatveev02,Kubala06}) and $\partial I_\text{E}/\partial V$. However, above the threshold line $V =\Delta$, indicated in red by (iv) in the schematic \Fig{fig:2}(b), a new \icot process $\mathcal{O}(\Gamma^2)$ \New{sets in: electrons tunnel onto and off
the} dot, while depositing an energy $\Delta$ \new{as sketched in \Fig{fig:0}(b).} This yields the characteristic step in the charge conductance~\cite{Heinrich04} in \Fig{fig:1}(a) at $V=\Delta$ all across the off-resonant regime~\cite{DeFranceschi01}. Our calculations show that the energy conductance $\partial I_\text{E}/\partial V$ also shows such an \icot feature at the corresponding line (iv) in \Fig{fig:1}(b). As expected, it changes sign when electron and hole processes change roles, similar to the sawtooth-shaped resonances discussed above, but now when tuning the level position through the \emph{center} of the off-resonant regime ($\varepsilon = -U/2$).
%A more radical difference appears in the magnitude of the inelastic step at $V=\Delta$ as function of the level position $\varepsilon$ in \Fig{fig:2}(a):
\New{Inspection of the magnitude of the inelastic step at $V=\Delta$ as function of the level position $\varepsilon$ in \Fig{fig:2}(a) reveals a dramatic difference:}
Whereas the charge conductance amplitude at $V=\Delta$ is smooth and featureless as $\varepsilon$ is varied, the energy conductance amplitude sharply drops \New{at (v)} when $\varepsilon\approx -\Delta$ or $\varepsilon+U \approx \Delta$. This big difference \New{also} shows up in \Fig{fig:1}(b) where the central part of the horizontal \icot onset is completely missing, in contrast to \Fig{fig:1}(a). The strong reduction of $\partial I_\text{E}/\partial V$ when entering the central region is remarkable: Everywhere in \Fig{fig:2}(a) we are still far from resonance, i.e., $|\varepsilon-\mu_{\text{L}}|, |\varepsilon+U- \mu_{\text{R}}| \gg T \gg \Gamma$. There is a second regime where the behavior of the energy conductance radically deviates from that of the charge conductance:
The bias dependence plotted in \Fig{fig:2}(c) shows at (vi) strong
\New{\emph{negative} differential energy conductance $\partial I_\text{E}/\partial V$ relative to the energy current $I_\text{E}$}
% \emph{negative} differential energy conductance $\partial I_\text{E}/\partial V$ for positive energy current,
by far exceeding the feature (v) discussed \New{below in magnitude.} As the inset indicates, after encountering the large energy current change at (v) the energy current drops \New{back} at (vi). \New{Also this} shows up in \Fig{fig:1}(b) as sharp blue (red) boundaries of the diamond-shaped region containing the label (vi) on the red (blue) background. \New{The} charge conductance in \Fig{fig:2}(c) is again featureless there. We now explain in three steps \New{(A)-(C)} how \New{physically these dramatic} differences come about, by following the vertical line in the schematic \Fig{fig:2}(b)
\New{at $\varepsilon=-0.45\,U$.}
%$\varepsilon=-0.55\,U$,
%i.e., in the central regime ($\Delta-U < \varepsilon < -\Delta$).
Our discussion explains which processes lead to changes in the occupations and transport rates~\cite{supplement} and were substantiated by numerical calculations.
%Our discussion of the processes below is substantiated by the calculation of the corresponding transition rates which appear naturally in the density-operator approach~\cite{supplement}.

(A) Starting in the state $\ket{\downarrow}$ at $V=0$ and increasing the bias we first hit the red threshold (iv) at $V=\Delta$. Beyond this line the excited state $\ket{\uparrow}$ \new{becomes} occupied by inelastic tunneling \new{[\Fig{fig:0}(b)]} and it relaxes by similar \New{inelastic} processes. \new{These transitions are} indicated by red arrows in the lower inset in \Fig{fig:2}(b). This gives an increased charge conductance as electrons find an additional path through the quantum dot while keeping the charge fixed to $N=1$, only virtually visiting other charge states $N=0,2$. In contrast, the energy current is relatively low due to two effects: 
First, real charge fluctuations are suppressed ($N=1$) and only \icot processes $\ket{\downarrow} \leftrightarrow \ket{\uparrow}$ occur, the rates for both of which are $\propto \Gamma^2$, much smaller than the rates $\mathcal{O}(\Gamma)$. 
Second, there is a significant partial cancellation of the energy currents of \New{these} $\mathcal{O}(\Gamma^2)$ processes, namely of a positive contribution due the \icot relaxation process $\ket{\uparrow} \to \ket{\downarrow}$, and a negative contribution due to the \icot excitation process $\ket{\downarrow} \to \ket{\uparrow}$ \new{[\Fig{fig:0}(b)]}.
This relates to the generic electron-hole symmetry between two \New{Coulomb-split SET resonances associated with filling a single orbital shell,} captured by our model~\cite{supplement}.

(B) \new{Increasing} $V$ further and crossing the green line (v) in \Fig{fig:2}(b), a two-step \set relaxation path is switched on, \new{indicated} by the green arrows in the \New{right} inset. This sharply increases the \New{magnitude of the} energy current at (v) in \New{\Fig{fig:2}(c)} since it lifts the above cancellation of \New{opposing} energy currents of comparable magnitude: the \icot relaxation process (red dashed downward arrow) is overridden by a much faster \New{$\mathcal{O}(\Gamma)$ two-step \set relaxation (green arrows),
$\ket{\uparrow} \to \ket{0}$ of \Fig{fig:0}(a) followed by $\ket{0} \to \ket{\downarrow}$.}
%$\ket{\uparrow} \to \ket{\uparrow\downarrow}$ followed by $\ket{\uparrow\downarrow} \to \ket{\downarrow}$.
This \New{composite} mechanism is called cotunneling assisted SET~\cite{Golovach04,Schleser05,Leijnse08a,Koller10} (COSET) and involves \new{\emph{real}} occupation of the \New{$N=0$ state, despite the prevalence of $N=1$ states due to Coulomb blockade.}

(C) When $V$ \New{finally} crosses the blue line (vi) in \Fig{fig:2}(b) the \emph{other} green relaxation path via \New{the $N=2$} state also becomes active as the upper inset \new{shows}.
Although still far from resonance, \emph{both} the $N=0$ \New{and} $N=2$ charge state become occupied for real \New{because both} $\mathcal{O}(\Gamma)$ relaxation pathways are turned on. Remarkably, this increased relaxation does not increase the energy current as above, but instead suppresses it. The reason is that the signed energy-current contributions from the two $\mathcal{O}(\Gamma)$ pathways now cancel each other\new{---in contrast to case (B)}
% the regime containing the label (v),
where only one such pathway is active\new{---}thereby strongly reducing the energy current.
%, similar to case (A) above.

\emph{Discussion.}---\New{We establish the complete classification of the nonlinear energy transport by noting the} additional feature due to the tunneling of \emph{pairs}~\cite{Leijnse09a} of either electrons or holes, which is again more prominent in \new{the energy conductance \new{[line (vii) in \Fig{fig:1}]}.}
%Our classification of $\mathcal{O}(\Gamma)$ plus $\mathcal{O}(\Gamma^2)$ effects is completed by noting \new{an} additional feature due to the tunneling of \emph{pairs}~\cite{Leijnse09a} of either electrons or holes which is again more prominent in \new{the energy conductance \new{[line (vii) in \Fig{fig:1}]}.}
%$\partial I_\text{E}/\partial V$ than in $\partial I_\text{C}/\partial V$.
Much of the above remains qualitatively the same when including a junction and spin\cite{Pustilnik00,Koenig05rev,Paaske10,Pletyukhov11} dependence of the tunneling constants $\Gamma_{\alpha\sigma}$, or a combined voltage ($\mu_{\text{L}} > \mu_{\text{R}}$) and thermal bias~\cite{Esposito09,Leijnse10,Wang12} ($\Gamma \ll T_{\text{L}} < T_{\text{R}} \ll  U$): Interestingly, in the latter case
\new{the \coset resonances may be used experimentally to
estimate the thermal gradient \emph{in situ}~\cite{supplement}.}

The above described nonequilibrium competition between real and virtual processes together with the sign of energy currents leads to an unexpectedly rich energy current spectrum. Importantly, for \New{more complex} multi-level quantum dots the above identified elementary signatures are just repeated every time a new electronic orbital is filled when scanning the gate voltage. Our model captures this generic pattern which is well attested experimentally for the charge current. However, we even find~\cite{supplement} that several replicas of these features can appear \New{in energy transport} inside the Coulomb blockade regime (e.g., as negative $\partial I_\text{E}/\partial V$ \New{relative to $I_\text{E}$}), but also \emph{outside}, at higher voltage, again in stark contrast to charge transport. Combining a three-terminal setup with measurements of the energy conductance may thus reveal 
% be a promising spectroscopy tool, providing
new qualitative information about relaxation processes. The much enhanced effect of \coset at (v) and (vi) in the energy transport of \Fig{fig:1}(b) should be experimentally accessible since even the weaker \coset features in the charge transport of \Fig{fig:1}(a) have  been measured~\cite{Schleser05,Sapmaz05b,Huettel09}.

Our results also indicate that the analysis of two-terminal thermoelectric measurements requires extra care due to the lack of gate-spectroscopic information. Often the two regimes of pure \icot [label (iv) in Fig.~\ref{fig:2}(b)] and \coset [label (v)] are not distinguished. For the charge conductance this may not \new{seem} so important, but our results show that for the energy conductance this distinction is absolutely vital. Theoretical descriptions used to model scanning-probe experiments and quantum dots in the Coulomb blockade regime are often based on effective models including only effective exchange and potential scattering terms. These may fail badly for the energy current since they include only \new{\icot} (keeping $N=1$ fixed) and eliminate the important real charge fluctuations (to $N=0$ and $2$) involved in COSET.
%In the present case
\New{In~\Fig{fig:1}(b)} such an approximation is suitable only in a limited regime, the triangle labeled (iv).
%. However,
Extending the model to include more inelastic excitations \New{further narrows it down~\cite{supplement}. Nonlinear energy transport
%, e.g., involving many vibrations (nano-electromechanical systems) or various spin excitations (spin caloritronics),
 thus requires} careful consideration: One needs \New{a} physical \emph{ model} allowing for charge fluctuations as well as a nonequilibrium transport theory that captures at least the first \emph{two} leading-orders of tunneling processes for strong interaction. Experimentally, even higher-order tunneling effects may be important~\cite{Svensson13} and it is of interest to explore
 renormalization effects~\cite{Paaske06,Pletyukhov12a,Kretinin12}.
 % on the energy current.

We thank C. Cuevas, \new{D. DiVincenzo}, F. Haupt, \new{M. Hell}, M. Leijnse, P. Reddy, and R. Saptsov for valuable discussions.
NMG, CBMH and DS thank the Institute for Theory of Statistical Physics, RWTH Aachen University, where substantial parts of this work have been performed. 
This work is part of the D-ITP consortium, an NWO program funded by the Dutch Ministry of Education, Culture and Science (OCW). 
CBMH and DS were supported by the DFG through the Emmy-Noether Program under SCHU 2333/2-1.

%%%%%%%%%%%%%%%%%%%%%%%%%%%%%%%%%%%%%%%%%%%%%%%%%%%%%%%%%
\newpage
\phantom{O}
\newpage
\onecolumngrid
\setcounter{figure}{0}

\begin{center}
\textbf{\large Supplement to ``Charge fluctuations in nonlinear heat transport"}
\end{center}

\section{Method: real-time density-operator transport theory}
%--------------------------------------------------
In this part of the supplementary material we provide the basic physical background as well the technical details of our density-operator approach to thermotransport through strongly interacting nanostructures. It combines the general reduced density-operator approach, accounting for the strong local interaction $U$, with a real-time diagrammatic method for the calculation of self-energies that determine transition rates of a master equation. The approach is perturbative and well-controlled in the regime where the tunneling rates are well below the electrode temperatures, $\Gamma\ll T_\text{L},T_\text{R}$. We go beyond standard approaches by including the competition of all $\mathcal{O}(\Gamma)$ and $\mathcal{O}(\Gamma^2)$ transition rates into the master equation for the state occupations and---what is new in particular---calculating the full energy-transport rate matrix to $\mathcal{O}(\Gamma^2)$. See Refs.~\onlinecite{supLeijnse08a,supSchoeller09a,supLeijnse09a,supKoller10,supSaptsov12a,supSaptsov14a} for more details and applications in the context of charge transport.

%%%%%%%%%%%%%%%%
\subsection{Master equation and current formulas---competition $\mathcal{O}(\Gamma) \leftrightarrow \mathcal{O}(\Gamma^2)$ for strongly interacting systems}
\label{sec:master-equation}
Before we present the detailed derivation of the theoretical method in the next section, we discuss the general form of the resulting master equation and current formulas and their physical significance.
In particular, we summarize the dependence of the transition rates on the level position and bias voltage in support of the central discussion in the main article explaining Fig.~3.

\begin{figure}[t]
\centering
\subfloat[]{\raisebox{-0.5\height}{
\includegraphics[height=0.12\textheight]{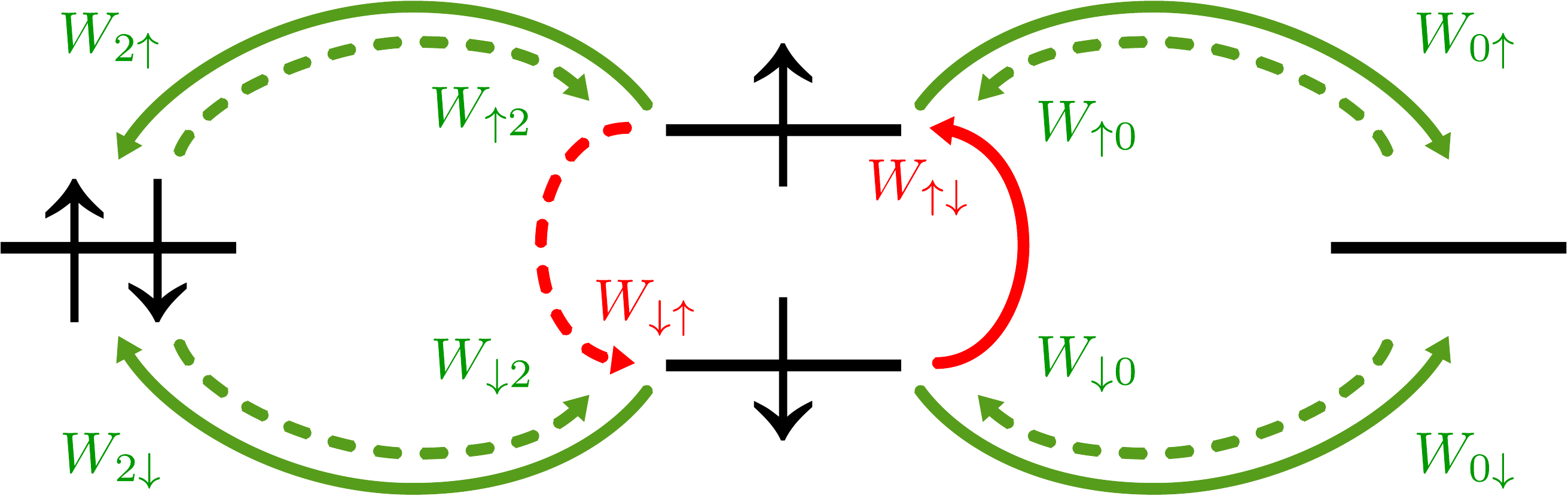}}
}
\subfloat[]{\raisebox{-0.5\height}{
\includegraphics[height=0.17\textheight]{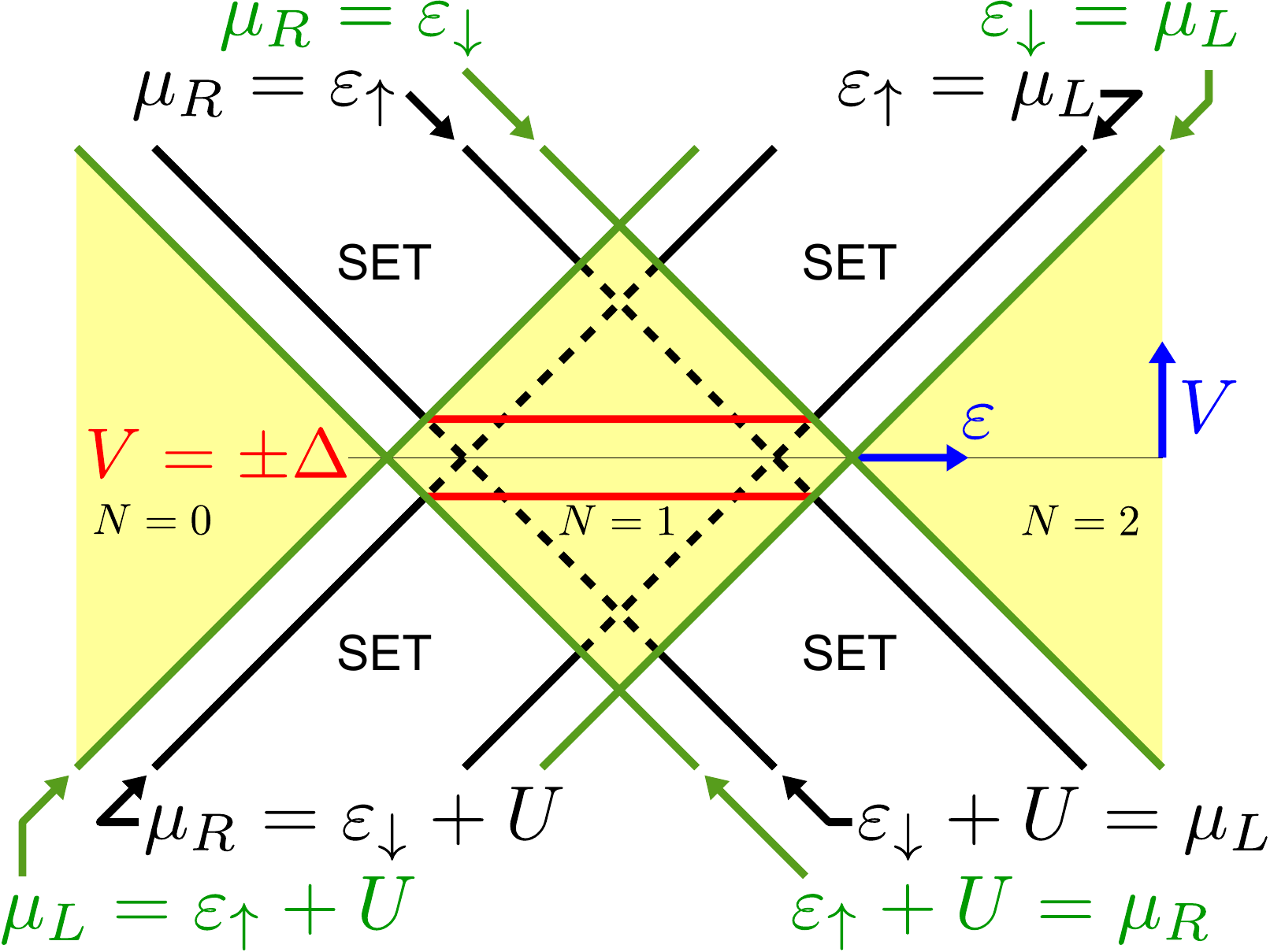}}
}
\protect\caption{(a) Transitions between the states of the Anderson quantum dot induced by tunneling processes in our $\mathcal{O}(\Gamma)$ plus $\mathcal{O}(\Gamma^2)$ approximation.
(b) Resonance lines for SET processes (green, black) and ICOT processes (red). COSET resonances occur at SET resonances that fall outside the SET regime (dashed black) but are still beyond the ICOT threshold (red), i.e., $|V|>\Delta$. The pair tunneling resonance is not indicated to maintain clarity, see text. The width and height of the central diamond equal $U+\Delta$.
\label{fig:transitions} }
\end{figure}

Our master equation describes the nonequilibrium state of the quantum dot, in particular the occupations $p_i$ of the quantum states $\ket{i}$, $i=0,\uparrow,\downarrow, 2$, in a statistical mixture $\rho = \sum_i p_i \ket{i}\bra{i}$  in the stationary long-time limit
\begin{align}
  \frac{d}{dt}
  \col{p_0}{p_\uparrow}{p_\downarrow}{p_2}
  \overset{t\rightarrow\infty}{=}
  \col{0}{0}{0}{0}
  =
  \begin{bmatrix}
    W_{00}         &    W_{0        \uparrow}    &    W_{0       \downarrow} &   W_{0        2} \\
    W_{\uparrow 0}   &    W_{\uparrow\uparrow}    &   W_{\uparrow\downarrow}   &  W_{\uparrow 2} \\
    W_{\downarrow 0} &    W_{\downarrow\uparrow}   &  W_{\downarrow\downarrow}  &  W_{\downarrow 2} \\
    W_{20}         &    W_{2          \uparrow} &    W_{2        \downarrow}  &  W_{2        2}
  \end{bmatrix}
  \col{p_0}{p_\uparrow}{p_\downarrow}{p_2}
  \label{eq:master-equation}
\end{align}
For clarity in the indexing of $p$ and $W$ we denoted the two-electron singlet by $\ket{2}:=\ket{\uparrow\downarrow}$. The transition rates obey an exact sum rule
$
  \sum_{i} W_{i j} = 0
$
which derives from the probability conservation $\sum_i p_i=1$. We can thus restrict our discussion of the \emph{transition rates} $W_{ij}$ to those between different states $i \neq j \in \{0,\uparrow,\downarrow,2\}$ as sketched in \Fig{fig:transitions}(a) in the same manner as in the main article [inset boxes to Fig.~3(b)]:

\begin{itemize}
\item
  \emph{Resonant/single-electron tunneling (SET) rates}
  $W_{0\sigma}$, $W_{\sigma 0}$, $W_{\sigma 2}$, $W_{2 \sigma}$ \emph{with} $\sigma=\uparrow,\downarrow$\\
  These rates are dominated by their $\mathcal{O}(\Gamma)$ contributions, which coincide with Fermi's Golden Rule expressions
  \begin{align}
    W_{\sigma 0}, W_{0\sigma} = \sum_\alpha \Gamma_{\alpha} f^\pm ((\varepsilon_\sigma-\mu_\alpha)/T_\alpha),
    \qquad
    W_{2\bar{\sigma}},W_{\bar{\sigma}2} = \sum_\alpha \Gamma_{\alpha} f^\pm ((\varepsilon_\sigma+U-\mu_\alpha)/T_\alpha),
  \end{align}
  where $f^\pm=(e^{\pm x}+1)^{-1}$ and $\bar{\sigma}=-\sigma$ and $\mu_\text{L/R}=\pm V/2$.
  Whenever one of the Fermi functions changes between 0 and 1, a transport resonance is naively expected [i.e., without solving the master equation~\eq{eq:master-equation}],
  i.e., at one of the eight lines defined in the $(\varepsilon,V)$ plane by
  \begin{align}
    \varepsilon_{\sigma}    = \mu_\alpha
    \quad
    \text{or}
    \quad
    \varepsilon_{\sigma} +U = \mu_\alpha
    \qquad
    \text{for $\alpha=L,R$ and $\sigma = \uparrow,\downarrow$.}
    \label{eq:set-resonance}
  \end{align}
  These are sketched in \Fig{fig:transitions}(b):
  in the indicated SET regimes transport by $\mathcal{O}(\Gamma)$ transitions between two ground states is energetically possible,
  whereas outside these regimes a single ground state is stable.
  The SET rates vary exponentially when going off-resonance, i.e., $|\varepsilon_\sigma-\mu_\alpha| \gg T$.
  These rates have $\mathcal{O}(\Gamma^2)$ corrections which describe the shift and broadening of the resonance.
  However, the naive expectation  \eq{eq:set-resonance} for a resonance is not precise in general, even when staying in $\mathcal{O}(\Gamma)$: In \Sec{sec:T-dependence} we will discuss anomalous temperature-dependent shifts [cf.~\Eq{eq:shift}] which can be understood only when actually solving the master-equation.
  Still, the naive resonance positions are useful as they provide a framework for discussing of the effects on which the main article focuses on.
\item
  \emph{Inelastic cotunneling (ICOT) rates}
  $W_{\uparrow\downarrow}$, $W_{\downarrow\uparrow}$\\
  These rates are due to $\mathcal{O}(\Gamma^2)$ processes and cause transitions between the $N=1$ spin states, see \Fig{fig:transitions}(a).
  The inelastic excitation rate $W_{\uparrow\downarrow}$ is exponentially suppressed for $|V|<\Delta$
  but starting from $|V|=\Delta$ it steadily increases as one goes beyond the red onset line in \Fig{fig:transitions}(b).
  This is directly reflected by the inelastic current as function of the bias plotted in the inset to Fig.~3(c) in the main article and also below in \Fig{fig:sup-4}(d).
  The inelastic relaxation rate $W_{\downarrow\uparrow}$, in contrast, is nonzero at zero bias and increases from there.
  \\
  The explanation of the central results of the main article in Fig.~3 revolves around the fact
  that at low bias the energy current contributions due to $W_{\uparrow\downarrow}$ and $W_{\downarrow\uparrow}$ cancel each other [regime (iv)].
  This cancellation is undone once the relaxation rate $W_{\downarrow\uparrow} = \mathcal{O}(\Gamma^2)$ gets ``overridden'' by \emph{either} the SET rate $W_{0\uparrow}$ \emph{or} $W_{2\uparrow}$ of $\mathcal{O}(\Gamma)$ is switched on [crossing either line (v)].
Once \emph{both $W_{0\uparrow}$ and $W_{2\uparrow}$} are switched on their energy contributions give rise to a new cancellation and the energy current is again reduced [regime (vi)], see \Fig{fig:transitions}(a).
\item
  \emph{Electron pair/hole pair tunneling rates}
  $W_{02}$ \emph{and} $W_{20}$\\
  These transition rates are often overlooked and were first pointed out in Ref.~\onlinecite{supLeijnse09a}.
  They are generated in $\mathcal{O}(\Gamma^2)$
  and lead to \emph{real} occupation of the charge state 2 (0)  in regimes where only
  charge states 0 and 1 (1 and 2) are accessible, respectively, by $\mathcal{O}(\Gamma)$ SET processes, see \Fig{fig:transitions}(a).
  This leads to a step resonance in $\partial I_\text{C}/\partial V$, well below the SET resonance~\cite{supLeijnse09a} [not indicated in \Fig{fig:transitions}(b) to maintain clarity, but see line (vii) in Fig. 2 of the main article]  that would lead to real occupation of this state when staying in $\mathcal{O}(\Gamma)$.
  Remarkably, in the energy conductance these pair-tunneling processes appear much more prominently, as mentioned in the discussion section of the main article.
\end{itemize}
An important aspect of our method is that we consistently calculate the \emph{nonequilibrium} populations of the quantum dot from the master equation~\eq{eq:master-equation}
instead of just combining some of the above transition rates with equilibrium occupations as is sometimes done in inelastic scattering theories.
In our discussion of the processes responsible for the features in Fig.~3 of the main article we have carefully verified our identification of the physics by comparing with the bias and energy-level dependence of the full numerically calculated rate matrices and the resulting occupations and current contributions.

Technically, the rates in the master equation \eq{eq:master-equation} can be determined from a superoperator kernel $W$ that we construct below [cf. \Eq{eq:perturbative-series} ff.] by letting it act on a state projector and taking diagonal matrix elements
\begin{align}
  W_{ij} = \bra{i} \Big[ W \ket{j}\bra{j} \Big] \ket{i}
  \label{eq:matrix-elements}
  .
\end{align}
The charge and energy currents flowing out of electrode $\alpha=L,R$ can be calculated from formulas with the physically appealing form of rate for transporting a quantity in a transition $i \to j$ $\times$ the occupation of the initial state $i$, summed over the initial ($i$) and final ($j$) states:
\begin{align}
  I_\text{C}^\alpha = \sum_{ij}(W_\text{C}^\alpha)_{ij} p_{j}
  ,
  \quad\quad
  I_\text{E}^\alpha = \sum_{ij}(W_\text{E}^\alpha)_{ij} p_{j}
  .
  \label{eq:currents}
\end{align}
Note that in the main article we have evaluated everything for $\alpha=\text{R}$ and dropped this superscript.

As we shall see in the next section, the calculation of the charge transport poses no real additional challenge since the required charge-current rate matrix $W_\text{C}^\alpha$ can be constructed from the reservoir-resolved contributions $W^\alpha$ required to calculate the transition matrix $W=\sum_\alpha W^\alpha$ for the master equation. This is due to the Kirchhoff law (charge conservation). For the heat or energy current no equivalent of the Kirchhoff law holds, and in particular when going to $\mathcal{O}(\Gamma^2)$ the energy-current rate matrix $W_\text{E}^\alpha$ needs to be calculated separately. This is the main technical achievement underlying the results reported in the main article.

%%%%%%%%%%%%%%%%
\subsection{Microscopic derivation of the master equation and the current formulas}
%%%%%%%%%%%%%%%%

In this section we derive the master equation \eq{eq:master-equation} and the current formulas \eq{eq:currents} starting from a general microscopic Hamiltonian $H_\text{tot}=H_{\text{res}}+H_{\text{d}}+H_{\text{tun}}$ where $H_{\text{res}}$ and $H_{\text{d}}$ act on the corresponding Hilbert spaces of the electronic reservoirs and the quantum dot, respectively. These two subsystems are coupled by $H_{\text{tun}}$ which is assumed to be bilinear in the electron field operators
and the reservoirs are effectively noninteracting, i.e., $H_{\text{res}}$ is quadratic in the fields.

%%%%%%%%%%%%%%%%
\subsubsection{General kinetic equation for the reduced density operator}

We first set up a perturbation theory in the tunnel coupling between the  quantum dot and the reservoirs. This is formulated from the start in Liouville-Fock space and has as a main advantage that one does not need to introduce an imaginary time or a Keldysh contour---\emph{even for non-equilibrium problems}: the time evolution of the density operator is fully described on a \emph{single, real time axis}. This comes at the price of having to deal efficiently with superoperators, which is achieved by introducing second quantization techniques directly in Liouville-Fock space, as we now briefly describe following Refs.~\onlinecite{supSaptsov12a,supSaptsov14a}, see also Ref.~\onlinecite{supSchoeller09a}.

\paragraph{Liouville-Fock space:}
Liouville-Fock space is the space spanned by the many-body operators acting on an underlying Hilbert-Fock space. It is convenient to introduce from the start multi-index notation $d_{1}$ for a field operator on the Hilbert-Fock space of the quantum dot, where $1=\left(\eta_{1},\sigma_{1}\right)$ and the extra index $\eta_{1}=+$ corresponds to a creation operator and $\eta_{1}=-$ to a destruction operator. Here $\sigma_1$ labels the internal quantum numbers of the quantum dot, which in our case is just the spin [for multi-level systems considered in Sec.~\ref{sec:genereric} below $\sigma_1$ also contains the orbital index]. To obtain field superoperators that generate the corresponding Liouville-Fock space we first consider the naive construction of field superoperators via the left action $\mathscr{G}_{1}^{+}\bullet=d_{1}\bullet$ and the right action $\mathscr{G}_{1}^{-}\bullet=\bullet d_{1}$ by a given field $d_1$ on an operator argument indicated by $\bullet$.
It turns out that this ansatz spoils the fermionic commutation rules. One can circumvent this problem\cite{supSaptsov12a} by taking linear combinations and involving the parity operator $(-{I})^{N}:=e^{i \pi N}$ associated with the electron number $N=\sum_\sigma d^\dag_\sigma d_\sigma$ by defining instead
\begin{equation}
\mathcal{G}_{1}^{q_{1}}\bullet=\frac{1}{\sqrt{2}}\left(d_{1}\bullet+q_{1}\left(-{I}\right)^{N}\bullet\left(-{I}\right)^{N}d_{1}\right), \qquad q\in\left\{ +,-\right\}.
\label{eq:G}
\end{equation}
The superoperators $\mathcal{G}_{1}^{q_{1}}$ indeed anticommute like fermionic fields: $\left[\mathcal{G}_{2}^{q_{2}},\mathcal{G}_{1}^{q_{1}}\right]_{+}=\delta_{q_{2},\overline{q}_{1}}\delta_{2,\overline{1}}\mathcal{I}$, where $\overline{1}=\left(-\eta_{1},\sigma_{1}\right)$, $\overline{q}=-q$ and $\mathcal{I}$ is the unit superoperator. The reservoir field superoperators $\mathcal{J}_1^q$ are defined analogously where now $1=\left(\eta_{1},\sigma_{1},\omega_{1}\right)$ and $\overline{1}$ again corresponds to the inversion of the $\eta$-index.
Assuming a grand-canonical density operator for the macroscopic reservoirs one finds that for odd $n$ $\left\langle \mathcal{J}_{n}^{q_{n}}\dots\mathcal{J}_{1}^{q_{1}}\right\rangle _{\text{res}}=0$, where $\braket{\bullet}_\text{res} = \tr_\text{res} \bullet \rho_\text{res}$, whereas for even $n$ one obtains a fermionic Wick theorem in Liouville-Fock space
\begin{equation}
\braket{ \mathcal{J}_{n}^{q_{n}}\dots\mathcal{J}_{1}^{q_{1}} }_{\text{res}}=\sum_{P}\left(-1\right)^{P}\prod_{\left\langle j,i\right\rangle }\braket{ \mathcal{J}_{j}^{q_{j}}\mathcal{J}_{i}^{q_{i}} }_{\text{res}},
\label{eq:Wick}
\end{equation}
where $P$ counts the number of permutations needed to join all pairs of contracted superoperators. The one-point correlation function is given by $\left\langle \mathcal{J}_{2}^{q_{2}}\mathcal{J}_{1}^{q_{1}}\right\rangle _{\text{res}}=\delta_{q_{2},-}\delta_{2,\overline{1}}\gamma_{1}^{q_{1}}$ with $\gamma_{1}^{+}=1$ and $\gamma_{1}^{-}=\tanh[\eta_{1}(\omega_{1}-\mu_{1})/(2T_{1})]$, which are up to a prefactor just the symmetric and antisymmetric parts of the Fermi-Dirac-distribution, respectively.

\paragraph{Perturbation theory in the tunnel coupling:}
We first follow the standard procedure of the density-operator approach by solving the Liouville-von-Neumann equation $\partial_{t}\rho_{\text{tot}}=-iL_\text{tot}\rho_{\text{tot}}$, where $L_{\text{tot}}\bullet=H_\text{tot} \bullet-\bullet H_\text{tot}$ is the Liouvillian of the total system. By Laplace transform its formal solution  $\rho_{\text{tot}}(t)=\exp(-iL_{\text{tot}}t)\rho_{\text{tot}}(t=0)$ becomes $\rho_{\text{tot}}(z)=-i(z-L_\text{tot})^{-1}\rho_{\text{tot}}(t=0)$.
Assuming the initial state factorizes as $\rho_{\text{tot}}(t=0) = \rho_{\text{res}}(t=0) \rho(t=0)$, where $\rho(t=0)$ denotes the initial density matrix of the dot, we can trace out the reservoir degrees of freedom.
Here the real-time approach deviates from traditional approaches by using the Liouville-space Wick theorem \eq{eq:Wick} \emph{right from the start}. From the resulting diagrammatic representation  one readily identifies a Dyson equation which propagates the initial density  operator \emph{of the quantum dot only}, $\rho(t=0)$,
\begin{equation}
\rho(z)=-\frac{i}{z-L_{\text{eff}}\left(z\right)}\rho(t=0),
\label{eq:rhoz}
\end{equation}
by an effective, energy-dependent Liouvillian
\begin{equation}
L_{\text{eff}}\left(z\right)=L+\sum_{k=0}^{\infty}\left.\tr_{\text{res}}\left(L_{\text{tun}}\frac{1}{z-\left(L+L_{\text{res}}\right)}\right)^{k}L_{\text{tun}}\,\rho_{\text{res}}\right|_{\text{irred.}}\text{.}
\label{eq:Leff-full}
\end{equation}
Here $L$ denotes the bare dot Liouvillian, $L\bullet=H_\text{d} \bullet-\bullet H_\text{d}$, which contains the energy scales $U$ and $\Delta$. We stress that we denote the density matrix of the dot and the dot Liouvillian by $\rho$ and $L$, respectively, i.e., we omit the subscripts d. With the mentioned assumptions this scheme is still exact so far, but we proceed by truncating this series. The superscript ``irred'' on the right hand side of \Eq{eq:Leff-full} indicates irreducibility of the expressions: this stipulates that after expanding in $L_{\text{tun}}$ one cannot move pairs of reservoir field superoperators apart from each other without changing any contraction-line crossing. The one-point correlation function in first nonvanishing order $\wick{}{\mathcal{J}_{2}^{q_{2}}}{}{\mathcal{J}_{1}^{q_{1}}}\mathcal{J}_{2}^{q_{2}}\mathcal{J}_{1}^{q_{1}}$ is obviously irreducible. For the second nonvanishing order, the pairs $\wick[1ex]{}{\mathcal{J}_{4}^{q_{4}}}{\mathcal{J}_{3}^{q_{3}}\mathcal{J}_{2}^{q_{2}}}{\mathcal{J}_{1}^{q_{1}}}\wick[1.5ex]{\mathcal{J}_{4}^{q_{4}}}{\mathcal{J}_{3}^{q_{3}}}{}{\mathcal{J}_{2}^{q_{2}}}\mathcal{J}_{4}^{q_{4}}\mathcal{J}_{3}^{q_{3}}\mathcal{J}_{2}^{q_{2}}\mathcal{J}_{1}^{q_{1}}$ and $\wick[1ex]{}{\mathcal{J}_{4}^{q_{4}}}{\mathcal{J}_{3}^{q_{3}}}{\mathcal{J}_{2}^{q_{2}}}\wick[1.5ex]{\mathcal{J}_{4}^{q_{4}}}{\mathcal{J}_{3}^{q_{3}}}{\mathcal{J}_{2}^{q_{2}}}{\mathcal{J}_{1}^{q_{1}}}\mathcal{J}_{4}^{q_{4}}\mathcal{J}_{3}^{q_{3}}\mathcal{J}_{2}^{q_{2}}\mathcal{J}_{1}^{q_{1}}$ are irreducible while $\wick{}{\mathcal{J}_{4}^{q_{4}}}{}{\mathcal{J}_{3}^{q_{3}}}\wick{\mathcal{J}_{4}^{q_{4}}\mathcal{J}_{3}^{q_{3}}}{\mathcal{J}_{2}^{q_{2}}}{}{\mathcal{J}_{1}^{q_{1}}}\mathcal{J}_{4}^{q_{4}}\mathcal{J}_{3}^{q_{3}}\mathcal{J}_{2}^{q_{2}}\mathcal{J}_{1}^{q_{1}}$ is reducible and must be omitted from \Eq{eq:Leff-full}. To perform the expansion we need the reservoir and tunnel Liouvillians: using the notation $\overline{\omega}_{1}=\eta_{1}\omega_{1}$ and a quite generic form of the tunnel and reservoir model we obtain
\begin{align}
H_{\text{tun}}^\alpha=\sum_{1}t_{1}\delta_{\alpha\alpha_1}\delta_{\eta+}d_{1}c_{\bar{1}}+\text{h.c.}
\qquad & \Rightarrow\qquad
L_{\text{tun}}^\alpha=\sum_{1}t_{1}\eta_{1}\delta_{\alpha\alpha_1}\sum_{q_{1}}\mathcal{G}_{1}^{\overline{q}_{1}}\mathcal{J}_{\overline{1}}^{q_{1}},\\
H_{\text{res}}=\sum_{1}\delta_{\eta+}\omega_{1}c_{1}c_{\bar{1}}
\qquad & \Rightarrow\qquad
L_{\text{res}}=\sum_{1}\overline{\omega}_{1}\mathcal{J}_{1}^{+}\mathcal{J}_{\overline{1}}^{-}.
\end{align}
Since we want to focus on the strong energy dependence introduced by the quantum dot on the energy currents, we use a flat density of states $\nu$ with some large cutoff $D$ (as compared to the energy scales $T, V, U, \Delta$ and $\varepsilon_\sigma-\mu_\alpha$).
This simplifies the present discussion and allows the energy integrations to be done analytically up to second order for $T_\text{L}=T_\text{R}=T$ in \Sec{sec:Gamma2},
but it presents no principal limitation of our method.
To be able to apply the Wick theorem \eq{eq:Wick} to \Eq{eq:Leff-full}, we need to commute the reservoir fields in $L_\text{tun}$ through the resolvents containing $L_\text{res}$.
Since these do not commute, $\left[L_{\text{res}},\mathcal{J}_{1}^{q}\right]_{-}=\overline{\omega}_{1}\mathcal{J}_{1}^{q}$, this changes the denominators of the intermediate propagators in \Eq{eq:Leff} and we obtain for the effective Liouvillian
\begin{align}
L_{\text{eff}}\left(z\right)=L+\sum_{k=2}^{\infty}\,\,\sum\!\!\!\!\!\!\!\!\!\!\!\!\int\limits _{\substack{1\dots k\\
q_{1}\dots q_{k}
}
}\nu^{k/2}t_{k}\dots t_{1}\left(\prod_{n=2}^{k}\mathcal{G}_{n}^{\overline{q}_{n}}\frac{1}{\sum_{m=1}^{n-1}\overline{\omega}_{m}+z-L}\right)\mathcal{G}_{1}^{\overline{q}_{1}}
\braket{
\left(\prod_{n=2}^{k}\mathcal{J}_{\overline{n}}^{q_{n}}\right)\mathcal{J}_{\overline{1}}^{q_{1}}
}_{\text{res}} ^{\text{irred.}}.
\label{eq:perturbative-series-full}
\end{align}
Here $\sum\!\!\!\!\!\!\int$ denotes a summation over reservoir and spin-indices, combined with integration over the energies $\overline{\omega_{i}}$.
The main challenge in going to higher orders---evenatually becoming prohibitive---is first the computation of the integrals over these energies and second the handling of the algebra of superoperators to efficiently evaluate the matrix elements \eq{eq:matrix-elements}.
However,  a rescaling of the integration variable $\overline{\omega_{1}}$ by a factor $1/T$ (to make the argument of the one-point correlation function dimensionless) reveals that $\Gamma/T\ll1$ is the relevant perturbation parameter, where $\Gamma$ denotes the scale of the tunnel coupling constants  $\Gamma_{1}=2\pi t_{1}^{2}\nu_{0}$.
Since in the main article we focus on the regime $\Gamma \ll T$ we truncate the Liouvillian at the second order. The effective Liouvillian in $\mathcal{O}(\Gamma^{1})+\mathcal{O}(\Gamma^{2})$ becomes
\begin{align}
L_{\text{eff}}\left(z\right) = & L+\Sigma(z)
\label{eq:Leff}
\\
\Sigma(z) = & \sum\!\!\!\!\!\!\!\!\!\int\limits 
_{1q_{1}}\frac{\Gamma_{1}}{2\pi}\mathcal{G}_{\overline{1}}^{+}\frac{q_{1}\gamma_{1}^{q_{1}}}{\overline{\omega}_{1}+z-L}\mathcal{G}_{1}^{\overline{q}_{1}}\nonumber \\
 & +\sum\!\!\!\!\!\!\!\!\!\!\!\!\int\limits _{12q_{1}q_{2}}\frac{\Gamma_{1}\Gamma_{2}}{\left(2\pi\right)^{2}}\left(\mathcal{G}_{\overline{1}}^{+}\frac{1}{\overline{\omega}_{1}+z-L}\mathcal{G}_{\overline{2}}^{+}-\mathcal{G}_{\overline{2}}^{+}\frac{1}{\overline{\omega}_{2}+z-L}\mathcal{G}_{\overline{1}}^{+}\right)\frac{\overline{q}_{2}\gamma_{2}^{\overline{q}_{2}}}{\sum\limits _{i=1,2}\overline{\omega}_{i}+z-L}\mathcal{G}_{2}^{q_{2}}\frac{\overline{q}_{1}\gamma_{1}^{\overline{q}_{1}}}{\overline{\omega}_{1}+z-L}\mathcal{G}_{1}^{q_{1}}
 \label{eq:perturbative-series}
\end{align}
To obtain a description of the inelastic transport in the Coulomb blockade regime that is even \emph{qualitatively correct}
it is crucial to keep the  \emph{first two leading orders}, as demonstrated in the main article.
It should be noted that the interaction $U$ and inelastic excitation energy $\Delta$ (Zeeman energy for a spin-flip) are treated non-perturbatively in this scheme, appearing in the denominators through $L$.

%%%%%%%%%%%%%%%%
\subsubsection{Stationary-state occupation probabilities and rate matrices}

\paragraph{Stationary-state rate matrix $W$:}
Taking the stationary limit of \Eq{eq:rhoz} using $\rho(t=\infty) =-i \lim_{z\to i0^{+}} z \rho(z)$ we obtain the stationary-state equation determining the density operator denoted by $\rho:=\rho(t=\infty)$
\begin{equation}
L_{\text{eff}}\left(i0^{+}\right)\rho=0.
\label{eq:stationary}
\end{equation}
We write Laplace transforms $F(z)=\int^\infty_0 d t e^{i z t} F(t)$ at $z=i0^{+}$ as $F(i0^{+})$, without the risk of confusing them with $F(t)$ evaluated at $t=0$.
Using the charge- and spin-symmetry of the Anderson model one proves~\cite{supSaptsov12a} that the probabilities $p_i=\bra{i}\rho\ket{i}$ decouple from all coherences $\bra{i}\rho\ket{j}$  with $i\neq j$.
The set of equations \Eq{eq:stationary} containing only the probabilities form a closed subset involving only the rates \eq{eq:matrix-elements} with $W=-i L_{\text{eff}}$
[the factor $i$ is important for consistency with the current rate expressions below].
We thus obtain the stationary master equation \eq{eq:master-equation} announced earlier.\cite{supfootnote}

By truncating the series \eqref{eq:Leff-full} to \eq{eq:perturbative-series} we compute the rates for all physical processes of orders $\mathcal{O}(\Gamma)+\mathcal{O}(\Gamma^2)$ consistently. However, one has to be aware that by solving \Eq{eq:master-equation} for the stationary density operator one generates an expression containing higher-order terms, which are however negligible for $\Gamma/T\ll1$, see Ref.~\onlinecite{supLeijnse08a} for a discussion. To check that the obtained features are consistently calculated, we changed $\Gamma$ and checked that the observed features scale consistently with orders to which the kernel is calculated. This important analysis is discussed in \Sec{sec:Scaling-analysis}.

\paragraph{Stationary-state charge and energy current rate matrices $W_C$ and $W_E$:}
To obtain the stationary charge current all relevant information is already encoded in the rate matrix $W$ computed for the stationary density operator above. To extract this information one uses charge conservation~\cite{supSaptsov12a} and obtains
\begin{equation}
I_{\text{C}}^{\alpha}(t)
=\braket{ \tfrac{d}{dt}N^{\alpha}_{\text{res}} }(t)
=\braket{\,i\left[H_\text{tun}^\alpha,N^{\alpha}_{\text{res}}\right]_-}(t)
=\braket{\, i\left[N,H_{\text{tun}}^{\alpha}\right]_{-} \,}(t)
=\underset{\text{dot}}{\tr} \left[ -\tfrac{1}{2} L^{N,+}  \left[ \underset{\text{res}}{\tr} \left(-iL_{\text{tun}}^{\alpha}\right) \rho_{\text{res}} \right] \rho(t) \right],
\end{equation}
where $\braket{\bullet}(t) = \tr_\text{dot} \tr_\text{res} \bullet \rho_\text{tot}(t)$, and we have used $\sum_\alpha N_\text{res}^\alpha+N=\text{const.}$.
The outer expression involves a simple action of $L^{N,+} \bullet = N \bullet + \bullet N$,
whereas in the complicated part involving the trace over the reservoirs one can identify the same perturbative expressions involved in \Eq{eq:Leff-full}.
Repeating the perturbative analysis described there one obtains for the \emph{stationary} particle current
\begin{equation}
I_{C}^{\alpha}
= \frac{i}{2}\,\underset{\text{dot}}{\tr}L^{N,+}\Sigma^{\alpha}\left(i0^{+}\right)\rho,
\label{eq:IC}
\end{equation}
where $\rho$ is the stationary state determined from \Eq{eq:stationary} and $\Sigma^{\alpha}\left(z\right)$ denotes the reservoir-resolved part of the effective Liouvillian $L_{\text{eff}}\left(z\right)=L+\sum_\alpha \Sigma^{\alpha}\left(z\right)$, i.e., 
$\Sigma^{\alpha}$ is given by the second part of \Eq{eq:Leff-full}
by replacing $L_\text{tun} \to L_\text{tun}^\alpha$ with reservoir index fixed to $\alpha$.
The corresponding explicit result is given by \Eq{eq:perturbative-series} when fixing the left-most reservoir index to $\alpha$.
In contrast, for the energy current one has to consider
\begin{equation}
I_{\text{E}}^{\alpha}(t)
=\braket{ \tfrac{d}{dt}H_{\text{res}}^{\alpha} }(t)
=\braket{\, i\left[H_{\text{tun}}^{\alpha},H_{\text{res}}^{\alpha}\right]_- \,}(t)
= \braket{ i\sum_{2}\delta_{\alpha_{2}\alpha}\omega_{2} t_{2} d_{2} c_{\overline{2}} }.
\end{equation}
This term does look similar to terms appearing in the perturbative series for the density operator evolution as before and one can express the \emph{stationary} energy current as
\begin{align}
I_{\text{E}}^{\alpha}
& =\underset{\text{dot}}{\tr} i \Sigma_{\text{E}}^{\alpha} \left(i0^{+}\right)\rho
.
\label{eq:IE}
\end{align}
Its explicit from is obtained by replacing the left-most $L_\text{tun}$ in \Eq{eq:Leff-full} by
\begin{align}
L^\alpha_\text{tun,E} & =\frac{1}{2}\left[I_{\text{E}}^{\alpha},\bullet\right]_{+}
=\frac{i}{2}\sum_{2}\delta_{\alpha_{2}\alpha}t_{2}\left(\omega_{2}-\mu_\alpha\right)\mathcal{G}_{2}^{q_{2}}\mathcal{J}_{\overline{2}}^{q_{2}}.
\end{align}
The result again looks similar to \Eq{eq:perturbative-series}.
However, the energy factor $\omega_{2}$ now complicates the energy integrals, and this new energy-current kernel requires an independent calculation.

As for the density-operator kernel, the charge- and spin-symmetry of the Anderson model imply that~\cite{supSaptsov12a} in the current formulas \Eq{eq:IC} and \Eq{eq:IE} only terms involving the probabilities $p_i=\bra{i}\rho\ket{i}$ from the density operator $\rho$ contribute with matrix elements of the type \eq{eq:matrix-elements} of the superoperators 
$W_\text{C}^\alpha = \tfrac{1}{2} i \, L^{N,+} \Sigma^\alpha (i0^{+})$
and
$W_\text{E}^\alpha = i \Sigma^\alpha_\text{E}(i0^{+})$.
As noted before, in the main article we have evaluated everything for $\alpha=\text{R}$ and dropped this superscript.
We thus obtain the current formulas announced in \Eq{eq:currents}.

In summary, we have to compute two kernels---the kernel $\Sigma(i0^{+})=\sum_\alpha \Sigma^\alpha(i0^{+})$ of the kinetic equation to determine $\rho$ and the energy-current kernel $\Sigma_\text{E}^\alpha(i0^{+})$; the charge current kernel requires no separate computation.
Moreover, because the density operator is self-adjoint and the currents are real, we only need the imaginary parts of both kernels $\Sigma$ and $\Sigma^\alpha_{\text{E}}$, 
which have much simpler energy integrals than their real parts. The subset of their matrix elements coupling probabilities defined by \Eq{eq:matrix-elements} determines the rate matrices $W$, $W_\text{C}$ and $W_\text{E}$ required in the master equation~\eq{eq:master-equation} and the current formulas~\eq{eq:currents}. This completes the derivation.

%%%%%%%%%%%%%%%%
\subsubsection{Numerical implementation}

\paragraph{$\mathcal{O}(\Gamma)$ results:}
To $\mathcal{O}(\Gamma)$ it is possible to give a compact general expression for the charge and energy current, valid also for tunneling rates with a dependence on the junction ($\alpha=L,R$) and the spin ($\sigma$) (but collinear spin-polarization axes of the electrodes).
We used the basis adapted to the symmetries of the problem from \Cite{supSaptsov12a}. The integrations in the self-energy $\Sigma$ are carried out immediately by inserting the Sokhotski-Plemelj relation $ 1/(x+i0) = -i\pi \delta(x) + P 1/x$, giving that all principal value parts cancel out. Solving the master equation \eq{eq:master-equation} and inserting the stationary occupations into the current formula \Eq{eq:currents} we obtain for the charge current
\begin{equation}
I_{\text{C}}^{\alpha} =\frac{1}{2}\left(\begin{array}{cc}
1 & 1\end{array}\right)\left[\psi^{\alpha}-\xi^{\alpha}\xi^{-1}\psi\right],
\end{equation}
where
\begin{align}
\psi^{\alpha} & =\left(\begin{array}{c}
\frac{1}{2}\Gamma_{\uparrow\alpha}[\tanh\left(\frac{\varepsilon+U+\Delta/2-\mu_{\alpha}}{2T_{\alpha}}\right)+\tanh\left(\frac{\varepsilon+\Delta/2-\mu_{\alpha}}{2T_{\alpha}}\right)]\\
\frac{1}{2}\Gamma_{\downarrow\alpha}[\tanh\left(\frac{\varepsilon+U-\Delta/2-\mu_{\alpha}}{2T_{\alpha}}\right)+\tanh\left(\frac{\varepsilon-\Delta/2-\mu_{\alpha}}{2T_{\alpha}}\right)]
\end{array}\right),\\
\xi^{\alpha} & =\left(\begin{array}{cc}
\Gamma_{\uparrow\alpha} & \frac{1}{2}\Gamma_{\uparrow\alpha}[\tanh\left(\frac{\varepsilon+U+\Delta/2-\mu_{\alpha}}{2T_{\alpha}}\right)-\tanh\left(\frac{\varepsilon+\Delta/2-\mu_{\alpha}}{2T_{\alpha}}\right)]\\
\frac{1}{2}\Gamma_{\downarrow\alpha}[\tanh\left(\frac{\varepsilon+U-\Delta/2-\mu_{\alpha}}{2T_{\alpha}}\right)-\tanh\left(\frac{\varepsilon-\Delta/2-\mu_{\alpha}}{2T_{\alpha}}\right)] & \Gamma_{\downarrow\alpha}
\end{array}\right),
\end{align}
while $\psi=\sum_\alpha \psi_\alpha$ and $\xi=\sum_\alpha \xi_\alpha$.
For the energy current we obtain
\begin{equation}
I_{\text{E}}^{\alpha}=\frac{i}{2}
\left(\begin{array}{cc}
1 & 1\end{array}\right)
\Sigma_\text{E}^{\alpha}\left(\begin{array}{c}
1\\
(\zeta-\overline{\psi}\xi^{-1}\psi)/(8\Gamma)\\
-\xi^{-1}\psi
\end{array}\right),
\end{equation}
where we defined
\begin{align}
-i
\left(\begin{array}{cc}
1 & 1\end{array}\right)
\Sigma_\text{E}^\alpha= & \left(\begin{array}{c}
\sum_{\sigma}\frac{\Gamma_{\sigma\alpha}}{2}[\left(\varepsilon+U+\sigma\Delta/2\right)\tanh\left(\frac{\varepsilon+U+\sigma\Delta/2-\mu_{\alpha}}{2T_{\alpha}}\right)+\left(\varepsilon+\sigma\Delta/2-\mu_{\alpha}\right)\tanh\left(\frac{\varepsilon+\sigma\Delta/2-\mu_{\alpha}}{2T_{\alpha}}\right)]\\
\sum_{\sigma}\frac{1}{2}\Gamma_{\sigma\alpha}U\\
\frac{\Gamma_{\downarrow\alpha}}{2}[\left(\varepsilon+U-\Delta/2\right)\tanh\left(\frac{\varepsilon+U-\Delta/2-\mu_{\alpha}}{2T_{\alpha}}\right)-\left(\varepsilon-\frac{B}{2}\right)\tanh\left(\frac{\varepsilon-\Delta/2-\mu_{\alpha}}{2T_{\alpha}}\right)]+\Gamma_{\uparrow\alpha}\left(\varepsilon+\frac{U}{2}+\Delta/2\right)\\
\frac{\Gamma_{\uparrow\alpha}}{2}[\left(\varepsilon+U+\Delta/2\right)\tanh\left(\frac{\varepsilon+U+\Delta/2-\mu_{\alpha}}{2T_{\alpha}}\right)-\left(\varepsilon+\Delta/2\right)\tanh\left(\frac{\varepsilon+\Delta/2-\mu_{\alpha}}{2T_{\alpha}}\right)]+\Gamma_{\downarrow\alpha}\left(\varepsilon+\frac{U}{2}-\Delta/2\right)
\end{array}\right)^{T},\\
\zeta & =\sum_{\sigma\alpha}\frac{1}{2}\Gamma_{\downarrow\alpha}\left[
\tanh\left(\frac{\varepsilon+U-\Delta/2-\mu_{\alpha}}{2T_{\alpha}}\right)+\tanh\left(\frac{\varepsilon-\Delta/2-\mu_{\alpha}}{2T_{\alpha}}\right)
\right].
\end{align}
Furthermore we define with $\psi=\left(\begin{array}{cc} \psi_{\uparrow} & \psi_{\downarrow}\end{array}\right)^{T}$ a corresponding $\overline{\psi}=\left(\begin{array}{cc} \psi_{\downarrow} & \psi_{\uparrow}\end{array}\right)$, where $-\frac{1}{2}\xi^{-1}\psi$ corresponds to two entries of the right-most vector above.
%-------------------------
\paragraph{$\mathcal{O}(\Gamma^2)$ results:\label{sec:Gamma2}}
%-------------------------
When going to $\mathcal{O}(\Gamma)+ \mathcal{O}(\Gamma^2)$ a  number of things change:
(i) additional rates $\mathcal{O}(\Gamma^2)$ has to be computed and added to the $\mathcal{O}(\Gamma)$ to obtain $W$, $W_\text{C}$  and $W_\text{E}$.
(ii) the master equation \eq{eq:master-equation} has to be solved again for the occupations (with these new rates).
(iii) the currents need to be recomputed with the new rates and new occupations.
For the effects considered here, analytical formulas for the resulting currents do not bring much insight and the equations have been implemented numerically.
The main challenge lies in step (i) in the computation of the energy integrals listed below, after taking matrix elements.
We have numerically implemented these integrations for arbitrary temperatures $T_\text{L}$ and $T_\text{R}$, but at zero thermal bias $T_\text{L}=T_\text{R}=T$, the case focused on in the main article, these integrals can be calculated analytically. In the latter case, they can be expressed in the standard complex digamma function $\Psi$ and we list the results here.
For the integrals for the time-evolution (and charge current) kernel we obtain in agreement with Refs.~\onlinecite{supKubala06,supLeijnse08a}
\begin{align}
 &\frac{1}{2\pi}\text{Im}\, \int_{-D}^{D}\int_{-D}^{D}d\overline{\omega}_{1}d\overline{\omega}_{2}\frac{1}{\overline{\omega}_{1}+z_{3}}\frac{\gamma_{2}^{+}\gamma_{1}^{-}}{\sum\limits _{i=1,2}\overline{\omega}_{i}+z_{2}}\frac{1}{\overline{\omega}_{1}+z_{1}}=-\frac{\Psi\left(\frac{1}{2}-i\frac{z_{3}+\overline{\mu}_{1}}{2\pi T_{1}}\right)-\Psi\left(\frac{1}{2}-i\frac{z_{1}+\overline{\mu}_{1}}{2\pi T_{1}}\right)}{z_{3}-z_{1}},\\[2mm]
 &\frac{1}{2\pi}\text{Im}\, \int_{-D}^{D}\int_{-D}^{D}d\overline{\omega}_{1}d\overline{\omega}_{2}\frac{1}{\overline{\omega}_{1}+z_{3}}\frac{\gamma_{2}^{-}\gamma_{1}^{-}}{\sum\limits _{i=1,2}\overline{\omega}_{i}+z_{2}}\frac{1}{\overline{\omega}_{1}+z_{1}}\nonumber\\
&\quad= \coth\left(\frac{z_{2}+\overline{\mu}_{1}+\overline{\mu}_{2}}{2T}\right)\frac{\Psi\left(\frac{1}{2}-i\frac{z_{3}+\overline{\mu}_{1}}{2\pi T}\right)-\Psi\left(\frac{1}{2}-i\frac{z_{3}-z_{2}-\overline{\mu}_{2}}{2\pi T}\right)-\Psi\left(\frac{1}{2}-i\frac{z_{1}+\overline{\mu}_{1}}{2\pi T}\right)+\Psi\left(\frac{1}{2}-i\frac{z_{1}-z_{2}-\overline{\mu}_{2}}{2\pi T}\right)}{z_{3}-z_{1}}\nonumber\\
 &\quad\quad -\frac{\tanh\left(\frac{z_{3}+\overline{\mu}_{1}}{2T}\right)\phi\left(\frac{1}{2}-i\frac{z_{2}-z_{3}+\overline{\mu}_{2}}{2\pi T}\right)-\tanh\left(\frac{z_{1}+\overline{\mu}_{1}}{2T}\right)\phi\left(\frac{1}{2}-i\frac{z_{2}-z_{1}+\overline{\mu}_{2}}{2\pi T}\right)}{z_{3}-z_{1}},\\[2mm]
 &\frac{1}{2\pi}\text{Im}\,\int_{-D}^{D}\int_{-D}^{D}d\overline{\omega}_{1}d\overline{\omega}_{2}\frac{1}{\overline{\omega}_{2}+z_{3}}\frac{\gamma_{2}^{-}\gamma_{1}^{-}}{\sum\limits _{i=1,2}\overline{\omega}_{i}+z_{2}}\frac{1}{\overline{\omega}_{1}+z_{1}}\nonumber\\
&\quad= -\coth\left(\frac{z_{2}+\overline{\mu}_{1}+\overline{\mu}_{2}}{2T}\right)\frac{\Psi\left(\frac{1}{2}-i\frac{z_{3}+\overline{\mu}_{2}}{2\pi T}\right)-\Psi\left(\frac{1}{2}-i\frac{z_{3}-z_{2}-\overline{\mu}_{1}}{2\pi T}\right)+\Psi\left(\frac{1}{2}-i\frac{z_{1}+\overline{\mu}_{1}}{2\pi T}\right)-\Psi\left(\frac{1}{2}-i\frac{z_{1}-z_{2}-\overline{\mu}_{2}}{2\pi T}\right)}{z_{1}+z_{3}-z_{2}}\nonumber\\
 &\quad\quad -\tanh\left(\frac{z_{3}+\overline{\mu}_{2}}{2T}\right)\frac{\Psi\left(\frac{1}{2}-i\frac{z_{2}-z_{3}+\overline{\mu}_{1}}{2\pi T}\right)-\Psi\left(\frac{1}{2}-i\frac{z_{1}+\overline{\mu}_{1}}{2\pi T}\right)}{z_{1}+z_{3}-z_{2}}\nonumber\\
 &\quad\quad -\tanh\left(\frac{z_{1}+\overline{\mu}_{1}}{2T}\right)\frac{\Psi\left(\frac{1}{2}-i\frac{z_{2}-z_{1}+\overline{\mu}_{2}}{2\pi T}\right)-\Psi\left(\frac{1}{2}-i\frac{z_{3}+\overline{\mu}_{2}}{2\pi T}\right)}{z_{1}+z_{3}-z_{2}}.
\end{align}

The integrals for the energy current kernel have not been reported before. Defining $\phi\left(\frac{1}{2}-i\frac{z+\overline{\mu}_{1}}{2\pi T_{1}}\right)=-\Psi\left(\frac{1}{2}-i\frac{z+\overline{\mu}_{1}}{2\pi T_{1}}\right)+\ln\frac{D}{2\pi T_{1}}$ we obtain
\begin{eqnarray}
& &\frac{1}{2\pi}\text{Im}\, \int_{-D}^{D}\int_{-D}^{D}d\overline{\omega}_{1}d\overline{\omega}_{2}\frac{\overline{\omega}_{1}}{\overline{\omega}_{1}+z_{3}}\frac{\gamma_{2}^{+}\gamma_{1}^{-}}{\sum\limits _{i=1,2}\overline{\omega}_{i}+z_{2}}\frac{1}{\overline{\omega}_{1}+z_{1}}=-\ln\frac{D}{2\pi T_{1}}+\frac{z_{3}\Psi\left(\frac{1}{2}-i\frac{z_{3}+\overline{\mu}_{1}}{2\pi T_{1}}\right)-z_{1}\Psi\left(\frac{1}{2}-i\frac{z_{1}+\overline{\mu}_{1}}{2\pi T_{1}}\right)}{z_{3}-z_{1}},\\[2mm]
& &\frac{1}{2\pi}\text{Im}\, \int_{-D}^{D}\int_{-D}^{D}d\overline{\omega}_{1}d\overline{\omega}_{2}\frac{\overline{\omega}_{1}}{\overline{\omega}_{1}+z_{3}}\frac{\gamma_{2}^{-}\gamma_{1}^{-}}{\sum\limits _{i=1,2}\overline{\omega}_{i}+z_{2}}\frac{1}{\overline{\omega}_{1}+z_{1}}
\nonumber\\
&&\quad= -\coth\left(\tfrac{z_{2}+\overline{\mu}_{1}+\overline{\mu}_{2}}{2T}\right)
\text{Re}\frac{z_{3}\left[\Psi\left(\frac{1}{2}-i\frac{z_{3}+\overline{\mu}_{1}}{2\pi T}\right)-\Psi\left(\frac{1}{2}-i\frac{z_{3}-z_{2}-\overline{\mu}_{2}}{2\pi T}\right)\right]-z_{1}\left[\Psi\left(\frac{1}{2}-i\frac{z_{1}+\overline{\mu}_{1}}{2\pi T}\right)-\Psi\left(\frac{1}{2}-i\frac{z_{1}-z_{2}-\overline{\mu}_{2}}{2\pi T}\right)\right]}{z_{3}-z_{1}}
\nonumber\\
 &&\quad\quad +\frac{z_{3}\tanh\left(\frac{z_{3}+\overline{\mu}_{1}}{2T}\right)\phi\left(\frac{1}{2}-i\frac{z_{2}-z_{3}+\overline{\mu}_{2}}{2\pi T}\right)-z_{1}\tanh\left(\frac{z_{1}+\overline{\mu}_{1}}{2T}\right)\phi\left(\frac{1}{2}-i\frac{z_{2}-z_{1}+\overline{\mu}_{2}}{2\pi T}\right)}{z_{3}-z_{1}},\\[2mm]
 &&\frac{1}{2\pi}\text{Im}\, \int_{-D}^{D}\int_{-D}^{D}d\overline{\omega}_{1}d\overline{\omega}_{2}\frac{\overline{\omega}_{2}}{\overline{\omega}_{2}+z_{3}}\frac{\gamma_{2}^{+}\gamma_{1}^{-}}{\sum\limits _{i=1,2}\overline{\omega}_{i}+z_{2}}\frac{1}{\overline{\omega}_{1}+z_{1}}=-\phi\left(\frac{1}{2}-i\frac{z_{1}}{2\pi T_{1}}\right),
\end{eqnarray}
\begin{eqnarray}
 &&\frac{1}{2\pi}\text{Im}\, \int_{-D}^{D}\int_{-D}^{D}d\overline{\omega}_{1}d\overline{\omega}_{2}\frac{\overline{\omega}_{2}}{\overline{\omega}_{2}+z_{3}}\frac{\gamma_{2}^{-}\gamma_{1}^{-}}{\sum\limits _{i=1,2}\overline{\omega}_{i}+z_{2}}\frac{1}{\overline{\omega}_{1}+z_{1}}\nonumber\\
&&\quad= \coth\left(\tfrac{z_{2}+\overline{\mu}_{1}+\overline{\mu}_{2}}{2T}\right)
\text{Re}\frac{z_{3}\left[\Psi\left(\frac{1}{2}-i\frac{z_{3}+\overline{\mu}_{2}}{2\pi T}\right)-\Psi\left(\frac{1}{2}-i\frac{z_{3}-z_{2}-\overline{\mu}_{1}}{2\pi T}\right)\right]-\left(z_{1}-z_{2}\right)\left[\Psi\left(\frac{1}{2}-i\frac{z_{1}+\overline{\mu}_{1}}{2\pi T}\right)-\Psi\left(\frac{1}{2}-i\frac{z_{1}-z_{2}-\overline{\mu}_{2}}{2\pi T}\right)\right]}{z_{1}+z_{3}-z_{2}}
\nonumber\\
 &&\quad\quad + z_{3}\tanh\left(\frac{z_{3}+\overline{\mu}_{2}}{2T}\right)\frac{\Psi\left(\frac{1}{2}-i\frac{z_{2}-z_{3}+\overline{\mu}_{1}}{2\pi T}\right)-\Psi\left(\frac{1}{2}-i\frac{z_{1}+\overline{\mu}_{1}}{2\pi T}\right)}{z_{1}+z_{3}-z_{2}}\nonumber\\
 &&\quad\quad -\tanh\left(\frac{z_{1}+\overline{\mu}_{1}}{2T}\right)\text{Re}\frac{\left(z_{2}-z_{1}\right)\phi\left(\frac{1}{2}-i\frac{z_{2}-z_{1}+\overline{\mu}_{2}}{2\pi T_{1}}\right)-z_{3}\phi\left(\frac{1}{2}-i\frac{z_{3}+\overline{\mu}_{2}}{2\pi T_{1}}\right)}{z_{1}+z_{3}-z_{2}}.
\end{eqnarray}

%--------------------------------------------------------------------
\subsection{Recovering the Landauer approach}
\label{sec:Landauer}
%--------------------------------------------------------------------

Effective single-particle descriptions have been used to analyze energy transport
\cite{supGalperin07,supLopezSanchez13,supWhitney13,supLee13,supZotti14}.
A prominent one is the Landauer approach, which in particular was used to explain the asymmetry of heat dissipation in quantum devices caused by the energy flow\cite{supLee13}.
A natural question is how our density-operator approach relates to this method and we provide two answers to this question:
First, in general for $U=0$ the density-operator approach \emph{without perturbative expansion} in $\Gamma$  exactly reduces to a Landauer formula for any $\Gamma$ and $T$.
We show how this resummation can be simply performed by \emph{extending} our equations for $U=0$ to include broadening nonperturbatively.
Second, explicitly setting $U=0$ in the perturbative equations for $\Gamma \ll T$ used in the present article explicitly reduces to Landauer formula \emph{in the corresponding limit} (i.e., expanded in $\Gamma$). These two answers are discussed subsequently.

\subsubsection{General relation to exact solution for $U=0$ and arbitrary $\Gamma$, $T$}

It has been known since the beginning of the real-time diagrammatic formulation of the density-operator transport theory that the exact solution, taking a Landauer form, is incorporated~\cite{supSchoeller97hab} when performing the complete summation of the diagrammatic series \eq{eq:perturbative-series-full} for the self-energy $\Sigma$ or rate matrix $W$ in powers of $\Gamma$. This seemed only explicitly doable for simple concrete examples such as the noninteracting resonant level model (without spin)~\cite{supSchoeller97hab,supSchoeller00b}. The fact that in the density-operator approach the $U=0$ limit is not easy to see \emph{in general} may seem puzzling at first, since for, e.g., Green's functions approaches, based on an expansion in the interaction $U$ but exact in $\Gamma$, this limit is trivially recovered. However, the density-operator approach is built around the picture of a locally strongly interacting system: it is entirely formulated in terms of many-body density operators and therefore carries a certain ``overhead'' in the $U=0$ limit.

In the recently developed~\cite{supSaptsov12a,supSaptsov14a} formulation of the same theory---used in this article---this noninteracting limit becomes entirely transparent by exploiting the field superoperators \eq{eq:G}. This works for any number of orbitals including spin and allowing for arbitrary tunnel coupling matrix elements (breaking any local orbital and spin symmetry). For this only the wide-band limit needs to be assumed for simplicity, which prevails anyhow in practical applications and which is not an essential restriction. This allows for a single renormalization step~\cite{supSchoeller09a} to be performed with the following very simple result:
Starting from our series \eq{eq:perturbative-series-full}
\begin{enumerate}
\item
  One replaces all Liouvillians $L$ by $L -i \sum_1 \tfrac{1}{2} \Gamma_1 \mathcal{G}_{1}^{+} \mathcal{G}_{\bar{1}}^{-}$,
  which introduces an energy-independent level broadening. Importantly, this is \emph{derived}, not put in by hand.
\item
  One leaves out all terms in the series containing $\mathcal{G}_{1}^{-}$ superoperators since they are included via the first step.
\end{enumerate}
One obtains a new perturbation theory of the same form which is however much simpler. Importantly, taking this renormalized perturbation theory to finite order $2$ ($2N$) for a 2 ($N$) level quantum dot one recovers the \emph{exact} density-operator evolution for $U=0$. Moreover, for one-particle observables such as the charge current a first order calculation is already exact for $U=0$.

Thus, the $\mathcal{O}(\Gamma)+\mathcal{O}(\Gamma^2)$ perturbation theory~\eq{eq:perturbative-series} used in the present article already has the correct form of the exact solution, except that the bare Liouvillian is used (i.e., step 1 is missing). Since we focus on the limit $\Gamma \ll T$ we instead incorporate the corresponding term with $\mathcal{G}^{-}_1$ superoperators (step 2), which includes the leading broadening corrections, see also below. We thus neglect only the nonperturbative effect of broadening. For a detailed exposition and explanation of the physics underlying this formulation see~\Cite{supSaptsov12a} [cf. Eq. (79) there] and also Ref.~\onlinecite{supSaptsov14a}. Finally, we note that going beyond the perturbative limit in $\Gamma$ \emph{for finite} $U$ is possible within the framework that we use and seamlessly connects to the real-time renormalization-group approach~\cite{supSchoeller09a}. Such calculations are, however, very involved and for the Anderson model this has only recently been worked out for charge transport~\cite{supSaptsov12a}.

\subsubsection{Concrete recovery of the Landauer results for $U=0$ and $\Gamma \ll T$\label{sec:concrete}}

\begin{figure}[t]
\subfloat[]{\includegraphics[height=0.3\textheight]{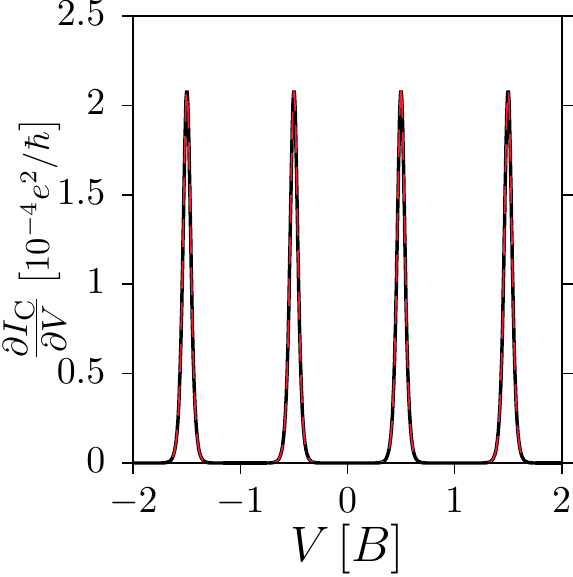}} \quad
\subfloat[]{\includegraphics[height=0.3\textheight]{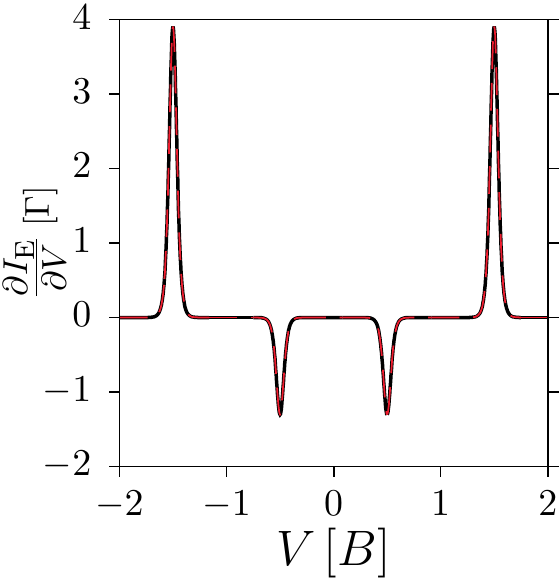}}
\protect\caption{ Comparison for the results for (a) the charge and (b) the energy conductance obtained with the reduced density-operator approach up to $\mathcal{O}(\Gamma)$ (red) and the Landauer formula\cite{supLee13,supZotti14} (black) for $U=0, \,\Delta=4\varepsilon\approx 83.3\,T,\, \Gamma=\frac{1}{3}\cdot 10^{-2}\,T$ and symmetric tunnel couplings.\label{fig:sup-1}}
\end{figure}
To concretely illustrate the above general conclusion we now simply insert $U=0$ into our equations,
first restricting attention to the $\mathcal{O}(\Gamma)$ contributions.
We then obtain the Landauer result for the charge and energy currents, respectively ($f$ notes the Fermi-Dirac-distribution)
\begin{align}
I_{\text{C}}  =\frac{1}{\pi}\int d\omega\, T\left(\omega\right)\cdot\left[f\left(\omega-\mu_\text{L} \right)-f\left(\omega-\mu_\text{R}\right)\right],
\quad\quad
I_{\text{E}}  =\frac{1}{\pi}\int d\omega\, T\left(\omega\right)\cdot\omega\cdot\left[f\left(\omega-\mu_\text{L} \right)-f\left(\omega-\mu_\text{R} \right)\right],
\end{align}
with the transmission function
\begin{equation}
T\left(\omega\right)=2\pi\sum_{\sigma}\frac{\Gamma_{\sigma L}\Gamma_{\sigma R}}{\Gamma_{\sigma L}+\Gamma_{\sigma R}}\delta\left(\omega-\varepsilon_\sigma\right)\,\text{.}
\end{equation}
Since we are in the leading order in $\Gamma/T$, the transmission function is a $\delta$-function of the energy $\omega$.
Including just the $\mathcal{O}(\Gamma^2)$ corrections in our density-operator approach (while maintaining $U=0$) only leads to broadening corrections.
This is done numerically in \Fig{fig:sup-1} and the corrections are negligibly small as expected for $\Gamma \ll T$.

Performing the above mentioned renormalization step one obtains the full $\Gamma$-broadened Lorentzian function of the Landauer result valid for any relation between~\cite{supSaptsov12a,supSaptsov14a} $\Gamma$ and $T$. However, we emphasize that even in that case there are \emph{no} inelastic cotunneling signatures in the transport since $U=0$. This shows the importance of our density-operator approach which is capable of capturing these processes and which were found to be especially important for the energy transport. Physically speaking, when not accounting for Coulomb blockade in the off-resonant transport regime the system is not in a particular well-defined charge state with well-defined inelastic excitation, which can thus not appear as a gate-voltage independent excitation.

%-------------------------
\section{Supporting results}
%-------------------------

In this second part of the supporting information we present additional results and discussion addressing specific issues raised in the main article.

%-------------------------
\subsection{Scaling with parameters---order of magnitude of charge, energy and heat current\label{sec:Scaling-analysis}}
%-------------------------

\begin{figure}[t]
\subfloat[]{\includegraphics[height=0.3\textheight]{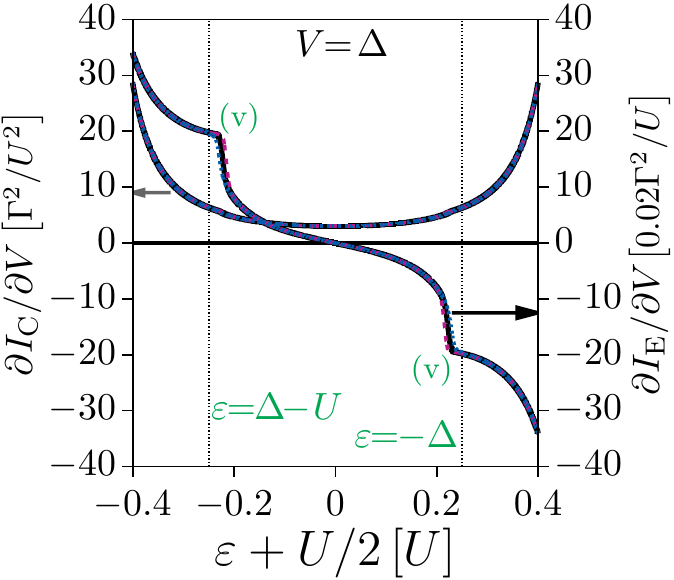}}\quad
\subfloat[]{\includegraphics[height=0.3\textheight]{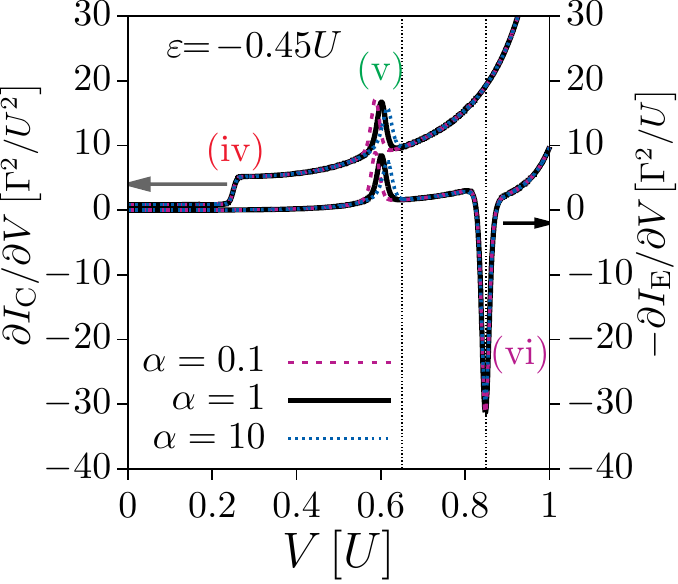}}
\\
\subfloat[]{\includegraphics[height=0.3\textheight]{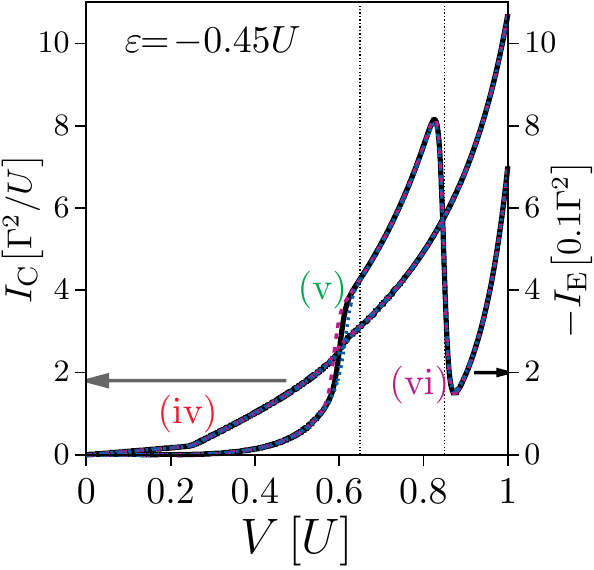}}\quad
\subfloat[]{\includegraphics[height=0.3\textheight]{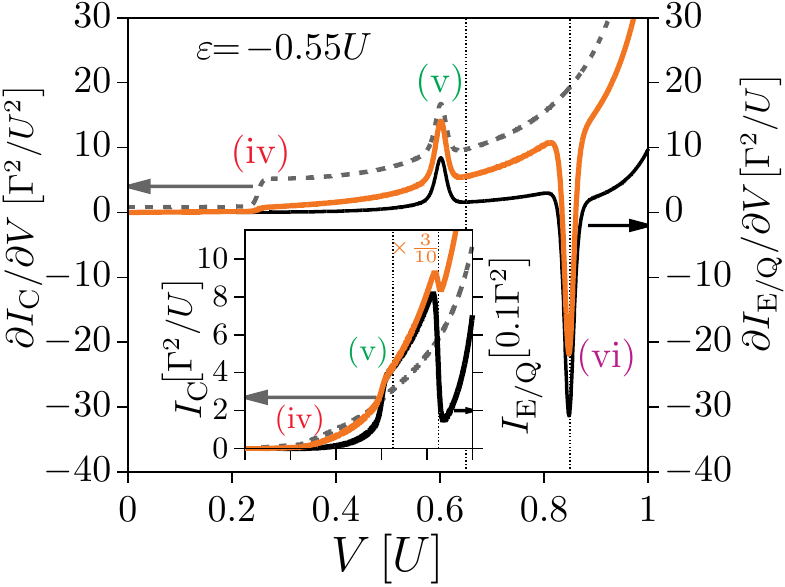}}
\protect\caption{(a) and (b): Scaling behavior of \emph{conductances} with respect to different $\Gamma$-values of Fig.~3(a) and (c) of the main article, respectively.
By changing $\Gamma\to\alpha\Gamma$ the results are rescaled according to $dI_{\text{C}}/dV\propto\Gamma^{2}$ (cf. Ref.~\onlinecite{supLeijnse09a}) and $dI_{\text{E}}/dV\propto\Gamma^{2}$.
(c): Scaling of the \emph{currents} corresponding to (b), shown in the inset to Fig.~3(c) of the main article.
The labels (iv)--(vi) are the same as in the main article. All sub-figures are for $\Delta=U/4\approx 83.3\,T,\, \Gamma=\frac{1}{3}\cdot 10^{-2}\,T$.
(d): Charge, energy and heat conductance for $\epsilon=-0.55U$, all other parameters are as in the other sub-figures. Inset: Corresponding currents. The full heat current $I_\text{Q}$ (solid orange line) is rescaled by 0.3 (independent of $\Gamma/U$) to fit into the plot.
\label{fig:sup-4} }
\end{figure}

In the main article we mentioned that the effects in the currents that we focus on---the \icot and \coset features in the Coulomb blockade---scale as (cf. also \Cite{supLeijnse09a}):
\begin{align}
  \partial I_\text{C}/\partial V \propto \Gamma^2/U^2,
  \quad \quad
  \partial I_\text{E}/\partial V  \propto  \Gamma^2/U,
  \label{eq:scaling}
\end{align}
for the parameter regime of interest where $\Gamma\ll U$.
It is important that this refers to \emph{corresponding features}, i.e., at voltages and level positions expressed in units of the interaction $U$.
The prefactors depend on the relative position of the \icot onset $\Delta / U$, $V/U$, $\varepsilon/U$; for the temperature $T$ dependence see below.
To illustrate this, we plot in \Fig{fig:sup-4}(a) and (b) the same data as in Fig.~3 (a) and (c) of the main text, respectively, but now calculated for rescaled $\Gamma$-values.
Plotted in dimensionless units \eq{eq:scaling} as function of $\varepsilon/U$ and $V/U$ respectively, the data fall onto two single curves, confirming the respective \emph{conductance} scaling behavior \eq{eq:scaling}. We thus conclude that our perturbative treatment to $\mathcal{O}(\Gamma^2)$ is consistent.

As can be seen in \Fig{fig:sup-4}(a)-(b), the only exception to the conductance scaling is the position and amplitude of the first COSET resonance position shifts.
\Fig{fig:sup-4}(c) shows the corresponding scaling of the \emph{currents}. This reveals that the temperature dependence corresponds \emph{only} to a shift of the position of the resonance (v): one switches from low values of $I_\text{E}$ to high values, the curve before and after this switch being \emph{independent} of $T$. It is only the $V$-position of the transition between them that shifts with $T$.
For the case of the charge current this relates to a known effect that is not relevant for the scaling discussion and is addressed separately below in~\Sec{sec:T-dependence}. With this in mind we can thus conclude that the magnitudes of the currents scale as
\begin{align}
  I_\text{C} \propto \Gamma^2/U,
  \quad \quad
  I_\text{E} \propto  \Gamma^2,
  \label{eq:scaling-current}
\end{align}
for corresponding features (as defined above).
To estimate the orders of magnitude for the experimentally accessible \emph{heat current} $I_\text{Q} = I_\text{E}-\mu_\text{R} I_\text{C}$
[all quantities refer the right electrode as in the main article],
we note that the sharp spectroscopic features in $I_\text{Q}$ are due to the energy current $I_\text{E}$ whereas the background is due to the convective part $\mu_\text{R} I_\text{C}$.
Importantly, $I_\text{E}$ and $\mu_\text{R} I_\text{C}$ scale in the same way, their relative magnitude being controlled by the relative excitation energy $\Delta/U$.
We illustrate this in \Fig{fig:sup-4}(d): the energy current is able to cause pronounced negative differential \emph{heat conductance}, i.e., a notable sharp drop in the heat current magnitude at the second COSET resonance [marked by (vi) as in the main article].
We emphasize that from the experimentally measured heat current $I_\text{Q}$ one can obtain in a \emph{model-free way} the interesting energy current $I_\text{E}$
 by simply subtracting the known background formed by the convective contribution $\mu_\text{R} I_\text{C}$.

Thus, when estimating orders of magnitude we can ignore the difference between $I_\text{Q}$ and $I_\text{E}$.
Based on \Eq{eq:scaling-current} a first quick experimental estimate for the order of magnitude is obtained as
\begin{align}
I_\text{E}  \sim  U \,  I_\text{C}.
\label{eq:currents-relation}
\end{align}
Taking interaction energies $U \sim$~10$^2$-10$^3$~meV typical for molecular junctions and inelastic cotunneling currents $I_\text{C}\sim $~1-10~nA one obtains
$I_\text{E}\sim $~0.1-10~nW. For comparison, nowadays heat currents of 10s of nW can already be measured~\cite{supLee13}.
However, these measurements apply a larger bias $V^\text{SET}$ to achieve (nearly) resonant SET transport
$V^\text{SET} \geq U$  [such that $\mu_\text{L} > \varepsilon >  \mu_\text{R}$, cf. \Eq{eq:set-resonance}].
In this case we can estimate using the conductance quantum $e^2/h \approx 4\cdot 10^{-7} \Omega^{-1}$
\begin{align}
  I_\text{C}^\text{SET} \sim \frac{e^2}{h} \Gamma \sim 0.4 \left[ \frac{\Gamma}{\text{1 meV}}\right] \text{~nA},
  \quad  \quad  \quad
  I_\text{E}^\text{SET} \sim \frac{e^2}{h} \Gamma V^\text{SET} \sim  \left[ \frac{I_\text{C}^\text{SET}}{\text{1 nA}} \right] \cdot \left[ \frac{V^\text{SET}}{\text{1 meV}} \right] \text{~nW}.
\end{align}
To gain more insight into where the smaller heat current values in the inelastic transport regime come from, we now specify the  scaling of the heat dissipation for the inelastic energy current more precisely: The factor $U$ in  \Eq{eq:currents-relation} comes from the voltage being equal to the inelastic excitation $\Delta$ which is always a fraction of $U$ inside the Coulomb blockade regime [e.g., $V=\Delta=U/4$ at (iv) in Fig.~\ref{fig:sup-4}(c)]. Using \Eq{eq:scaling} the scaling of the inelastic heat current relative to that of the resonant case is found to be
\begin{align}
    I_\text{E}  \sim  \Delta \,  I_\text{C} \sim  \left( \frac{\Delta}{V^\text{SET}} \right) \, \left( \frac{\Gamma}{U} \right)  \, I_\text{E}^\text{SET}.
    \label{eq:icotresscaling}
\end{align}
The first suppression factor expresses that when staying in the Coulomb blockade regime one 
applies a smaller voltage than when going on resonance (leaving the Coulomb blockade regime), $V^\text{SET} \geq U > \Delta$, thus giving a smaller energy current.
The second factor is the effect of the Coulomb interaction $U > \Gamma$ suppressing the magnitude of the inelastic current relative to the resonant one.
Equation~\eq{eq:icotresscaling} indicates that for devices where $I_\text{E}^\text{SET} \sim$~10~$\mu$W one may expect the inelastic currents to be of the order of $I_\text{E}\sim$~10~nW,
which shows good prospects for coming within range of heat current measurements. We immediately emphasize that our estimates have been conservative in the following sense:
\begin{itemize}
\item
The above scaling is basically those around the center of the Coulomb blockade regime where all currents are minimal. Away from the center all the effects are enhanced.
\item
We used strictly perturbative formulas, assuming regimes where $\Gamma/U$ is small.
For larger values of $\Gamma$ and lower temperatures renormalization effects can greatly enhance the inelastic tunneling relative to the resonant tunneling, as is well known for charge transport, and expected also for energy transport, see \Sec{sec:nongeneric} for a simple formula.
\end{itemize}
We thus underestimate the heat currents that one should experimentally be able to resolve. A more detailed analysis is of interest but relies on details of a given experimental setup as well as goes beyond the scope of the present article and the methods employed here, which mainly serve to identify the basic processes of energy transport that were overlooked so far. As mentioned in the outlook of the main article, the second point is an interesting topic for future studies and requires a nonequilibrium renormalization group analysis to treat stronger tunnel couplings in the presence of interactions.

%------------------------
\subsection{Anomalous temperature dependence of COSET resonance positions\label{sec:T-dependence}}
%-------------------------
In Fig. 3(a) and (c) of the main article, as well as \Fig{fig:sup-4}, the first \coset resonance [labeled (v) as in the main article] do not appear exactly at the vertical dotted lines indicating the COSET resonance positions [those parts of the resonance lines \Eq{eq:set-resonance} that lie within the Coulomb blockade regime]. For the charge transport at the \coset resonance this effect has been first discussed in~\Cite{supGolovach04}. The mechanism causing the effect is however quite general and already plays a role in regimes where \emph{only SET transport occurs}. Understanding this mechanism is relevant for the COSET thermometry discussed in \Sec{sec:thermal-bias}.
In general, transport through a quantum dot shows a resonance whenever a new transition rate is ``switched on'' by changing the energy levels relative to the chemical potentials of the electrodes (either by gate or bias voltage or both). This changes the rate matrices $W$, $W_\text{C}^\alpha$, $W_\text{E}^\alpha$ causing the occupation probabilities to change [cf.~\Eq{eq:master-equation}] and adding a new contribution to the currents \eq{eq:currents}. Roughly speaking this happens when the relevant addition energy falls above or below the electrochemical potential of one of the electrodes. Although this is often ignored, the \emph{relevant criterion for resonance} is a significant change in the occupations and \emph{not this energy condition}. Already for the symmetric Anderson model there is a difference between the two conditions~\cite{supBonet02}.

The idea is most clearly understood from a simple case~\cite{supRomeike06b}: assume that a rate $W=\gamma f(\varepsilon)$ is being ``switched on'' by tuning an energy level $\varepsilon$ through $\mu$ of some electrode in the function $f(\varepsilon)=(e^{(\varepsilon-\mu)/T}+1)^{-1}$. We lump together all other rates depopulating this state into $\gamma'$ and both $\gamma'$ and $\gamma$ are assumed roughly constant [i.e., much weaker $\varepsilon$ dependence than $f(\varepsilon)$].
Naively one expects the resonance position to be $\varepsilon=\mu$, but this ignores interaction effects that come in through the stationary master equation~\eq{eq:master-equation}.
In this case the stationary master equation contains a line of the form
$-\gamma' p + W P=0$ with solution $p = W/\gamma' P$. The occupation $p$ of our state becomes comparable to the occupation $P$ of the other state (already populated) ``feeding'' it when
$W/\gamma' = \gamma/\gamma'\,f(\varepsilon) \sim 1$.
When the tunneling rate constants $\gamma$ and $\gamma'$ are identical, this reduces to the naive resonance condition $\varepsilon = \mu$.
However, when the ``feeding'' rate is larger, $\gamma \gg \gamma'$, the ratio already reaches 1 while still in the tail of the Fermi-distribution, $W/\gamma' \approx  \gamma/\gamma' e^{-(\varepsilon-\mu)/T} \sim 1$. Solving this modified condition, the naive expectation is modified to
\begin{align}
  \varepsilon = \mu + T \ln\left(\frac{\gamma}{\gamma'}\right).
  \label{eq:shift}
\end{align}
Physically speaking, while still far from resonance tunneling with a relatively large constant rate $\gamma$ is sufficient to alter the occupations of a state that relaxes slowly ($\gamma' \ll \gamma$). The resonance position thus \emph{shifts linearly with temperature}. For significantly differing rates $\gamma$ and $\gamma'$ this shift may drastically alter the excitation spectrum from the expected ``bare'' spectrum, but even small factors are noticeable.
This fact must already be taken into account in charge transport spectroscopy even when only $\mathcal{O}(\Gamma)$ processes are taken into account
and in this case has been experimentally observed~\cite{supDeshmukh02} and modeled in detail in in \Cite{supBonet02}.

Returning to  \coset resonance of interest in the main article, both $\mathcal{O}(\Gamma)$ and $\mathcal{O}(\Gamma^2)$ processes are competing. The above sketched situation of having largely differing tunneling rates naturally arises~\cite{supGolovach04}. In this case one considers the \emph{depopulation} of the excited spin state $\uparrow$ by an $\mathcal{O}(\Gamma)$ process which is switching on while a much smaller populating rate $ W_{\downarrow\uparrow} =  \mathcal{O}(\Gamma^2)$ is already present. This however, leads to the same scenario and the logarithmic prefactor scales as $\ln(U/\Gamma)$ in this case, making temperature shifts significant.
This again underlines the importance of solving a full nonequilibrium master equation for the occupations including interaction effects, cf. \Sec{sec:master-equation}.

%-------------------------
\subsection{Energy transport under combined voltage and thermal bias---COSET thermometry\label{sec:thermal-bias}}
%-------------------------

\begin{figure}[t]
\subfloat[]{\includegraphics[width=0.48\columnwidth]{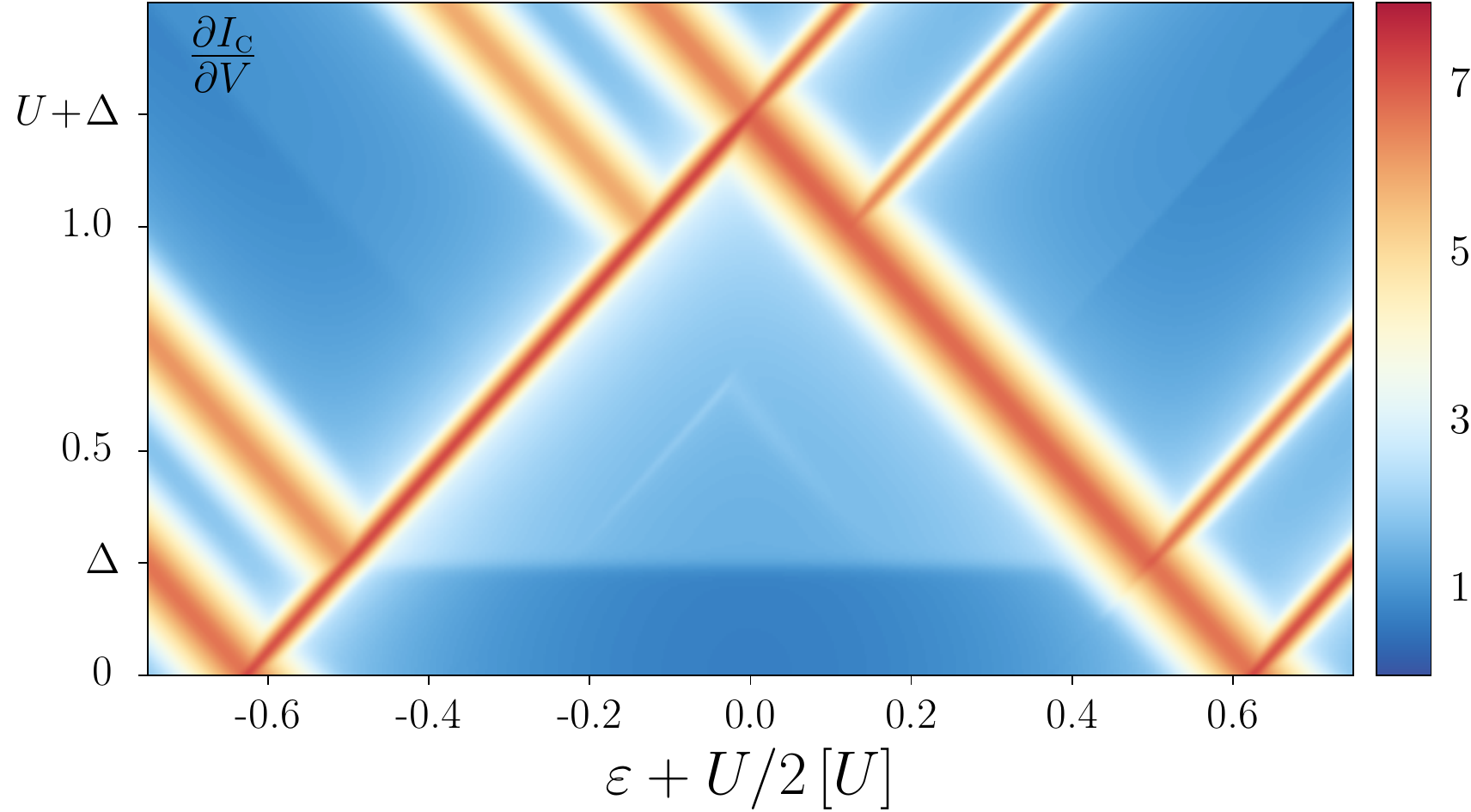}}\hfill
\subfloat[]{\includegraphics[width=0.48\columnwidth]{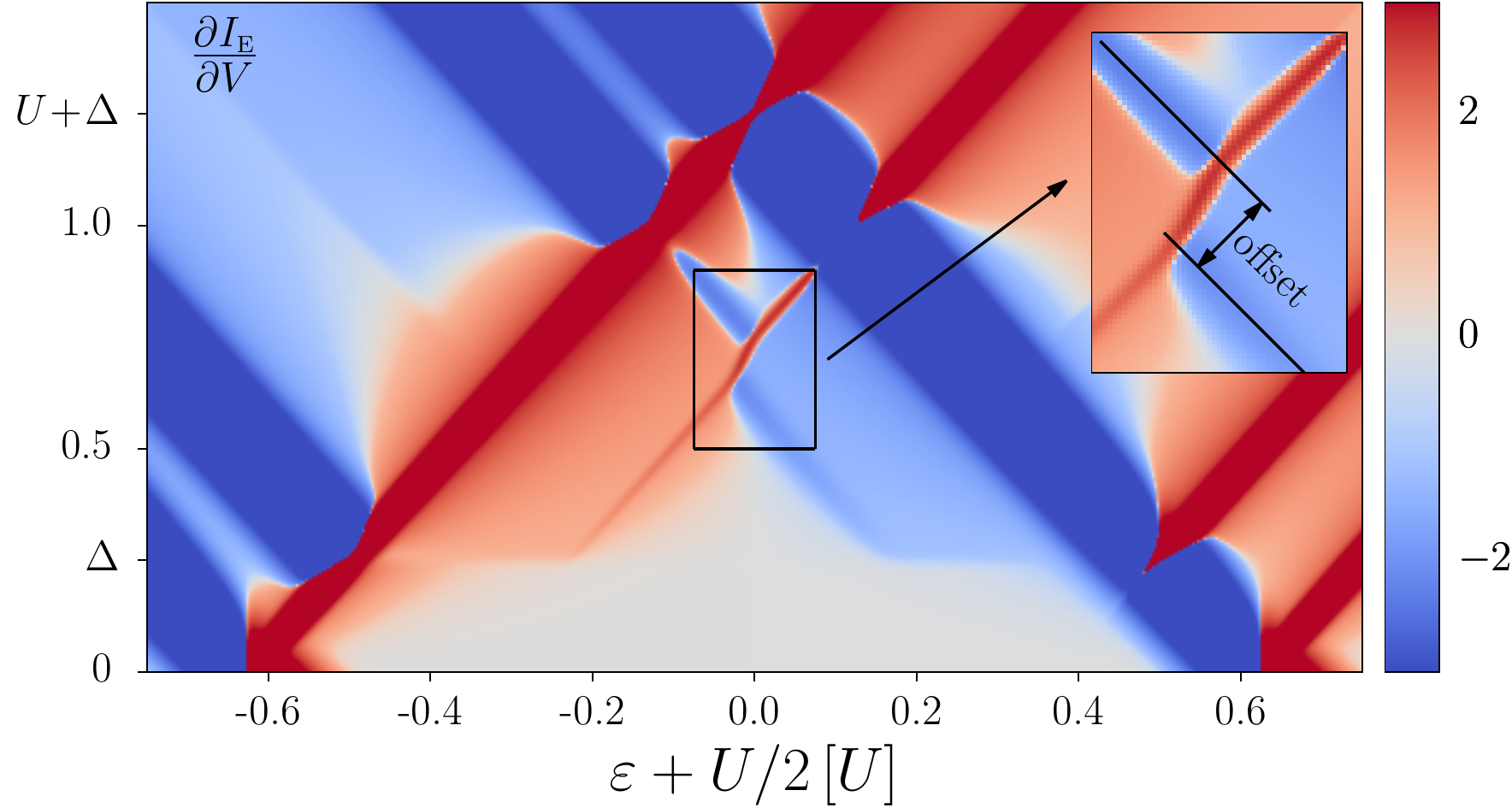}}
\protect\caption{Combined voltage and thermal bias plotted in the same way as in Fig. 2 of the main article. (a) Charge conductance  $\log_{10} \left( [\partial I_\text{C}/\partial V] / [\Gamma^2/U^2] \right)$. (b) Energy conductance $\text{slog}_{10}\left( [\partial (I_\text{E}/\partial V)] / [\Gamma^2/U] \right)$, where we use the signed log, $\text{slog}_{10}(x) := \sgn{x}\log_{10}(a|x|)$ for $a|x|>10$, which is linearized near zero, $\text{slog}_{10}(x):= ax/10$ for $a|x|<10$ using $a=20$. The inset magnifies the \coset lines whose offset can be used to accurately detect the temperature gradient. The parameters are 
$T_{\text{R}}=3T_{\text{L}}$ for $U=\frac{1}{3}\cdot\,10^3\, T_{\text{L}}$, $\Delta=U/4\approx 83.3\, T_{\text{L}}$ and $\Gamma=\frac{1}{3}\cdot 10^{-2}\,T_\text{L}$.
\label{fig:sup-3}}
\end{figure}
We briefly discuss the case of a joint voltage ($\mu_\text{L}>\mu_\text{R}$) and thermal bias ($T_\text{L}<T_\text{R}$) 
mentioned in the discussion section of the main article.
So far this has been studied in prior works\cite{supEsposito09,supLeijnse10,supWang12} only to $\mathcal{O}(\Gamma)$. The charge and energy conductance for a single-level Anderson dot are shown in \Fig{fig:sup-3}. Besides the expected asymmetry in the thermal broadening of the resonant tunneling lines associated with the different reservoirs, there are temperature-dependent offsets in both COSET lines that are highlighted in the frame in Fig.~\ref{fig:sup-3}(b).
These are caused by the mechanism discussed in the previous \Sec{sec:T-dependence}. In a thermoelectric setup this mechanism offers the interesting opportunity of \emph{converting a thermal bias into an energy splitting}.
Indeed, the largest offset between the blue \coset lines splits proportional to the temperature $T_\text{R}$ and can be used experimentally to detect a temperature gradient \textit{in situ}.
This can be done more directly and therefore more accurately than by taking the difference of resonant tunneling line broadenings alone. In this case, such \emph{COSET-thermometry} cannot be performed with the charge conductance.

%%%%%%%%%%%%%%%%
\subsection{Generic nature of the energy-current effects inferred from the Anderson model}\label{sec:genereric}
%%%%%%%%%%%%%%%%

In the main article we analyzed the nonlinear thermoelectric transport properties of a resonant level with strong Coulomb interaction and a well-defined spin-flip excitation $\Delta$ described by the Anderson model.
This suffices to classify nonlinear thermoelectric transport through an interacting nanoscale object, in the following sense:
these should generally be observable in a range of nanostructures with quasi-regular filling of well-separated electronic energy shells.
Here we first provide support for this conclusion from a general  point of view and then illustrate and further strengthen it with concrete multi-level calculations. We close by discussing where we expect deviations from the generic behavior.

\subsubsection{Beyond the Anderson model---general picture}

The general argument for the generic nature of the predicted effects is the following:
the features that we considered arose from the first two leading orders, $\mathcal{O}(\Gamma)$ and $\mathcal{O}(\Gamma^2)$,
that we took into account.
As explained in \Sec{sec:master-equation}, the rate matrices in the master equation due to these processes also couple charge states differing by \emph{two} electrons by pair tunneling processes, see \Sec{sec:master-equation}. However, these are typically not important in the Coulomb blockade regime on which we focus, but instead lead to qualitatively new features in the SET regime between subsequent resonances~\cite{supLeijnse08a}. With this insight, it is clear that the features depend mostly on neighboring charge states and will be repeated for every next orbital state being filled. This is strongly supported by experimental data on quantum dot systems for charge transport. There is no general reason why for energy currents this argument does not apply and the concrete multi-level calculations in the next section bear this out. In fact they show that the features in cases may proliferate rather than simply replicate.

\subsubsection{Example multi-level quantum dots---replication and proliferation of energy-current effects}

We now present results for a more complex model $H_\text{tot}=H_\text{d}+H_\text{res}+H_\text{tun}$, where
$H_\text{d}$ describes a quantum dot with \emph{two orbitals}:
\begin{equation}
H_{\text{d}}=
\sum_{i=1}^{2}\left[\sum_{\sigma}\left(\varepsilon_{i}+ \sigma \tfrac{1}{2} \Delta\right)d_{i,\sigma}^{\dagger}d_{i,\sigma}
+U N_{i} (N_{i} - 1)/2
\right]
 + U' N_1 N_2,
\label{eq:2levelQD}
\end{equation}
where we sum over $\sigma = \pm$ corresponding to $\uparrow,\downarrow$
and $N_{i} =\sum_\sigma d_{i,\sigma}^{\dagger}d_{i,\sigma}$ is the occupation operator for orbital $i=1,2$.
The junction Hamiltonian is generalized to include an amplitude for tunneling between each dot and each electrode $H_\text{tun}=  \sum_i t_i\sum_{k\alpha\sigma} (c_{k\alpha\sigma}^{\dagger}d_{i,\sigma}+\text{h.c.})$ but with otherwise the same assumptions as in the main article.
Also we again focus on the voltage-bias energy currents, i.e., $T_\text{L}=T_\text{R}=T \gg \Gamma$.
This model accommodates for two qualitatively new details that are experimentally relevant and well-known from charge transport studies:
\\
First, inelastic excitations can now be of two types:
\begin{itemize}
\item \emph{Spin splitting} $\Delta=\varepsilon_{1 \uparrow}-\varepsilon_{1 \downarrow} =\varepsilon_{2 \uparrow}-\varepsilon_{2 \downarrow}$:
as before, the dots are assumed to have {spin splittings} and these are assumed the same.
\item \emph{Orbital splitting} $\Delta'= \varepsilon_{2}-\varepsilon_{1}$: the {orbital splitting} is a new feature of the multi-orbital models.
\end{itemize}
Thus $\varepsilon_{1\sigma} = \varepsilon + \sigma \Delta/2$ as before and
$\varepsilon_{2\sigma} = \varepsilon + \sigma \Delta/2 +\Delta'$ for $\sigma=\uparrow,\downarrow$.
We assume that as one varies the gate voltage, $\varepsilon$ linearly varies while the orbital splitting $\Delta'$ remains constant.
\\
Second, the shell filling pattern in this model can be more complicated due to the presence of an additional \emph{interdot interaction} $U' < U$ and the orbital splittings $\Delta'$.
\begin{itemize}
\item 
For sufficiently large orbital splitting $\Delta'$ we first fill orbital 1 and then orbital 2. Once orbital 1 is filled, it provides a ``gating'' of orbital 2 by the interdot interaction $U'$, shifting the latter's addition energies.
In this case we have a very precise  electron-hole symmetry between the two Coulomb-split resonances associated with the filling each orbital.
\item
In contrast, for lower $\Delta'$ we may distribute electrons over the two orbitals (since $U'<U$) and then fill them up completely.
In this case the electron-hole symmetry is not exact but only approximate.
\end{itemize}
Before we discuss the example calculations, we formulate the key questions they should answer about the generic nature of the predictions in the main article.
For the Anderson model we identified two pronounced effects in the nonlinear energy conductance:
\begin{enumerate}[label=(\Alph*)]
\item\label{triangle}
\emph{The suppression triangle at low bias} above the onset of an inelastic excitation [labeled (iv) in the main article].\\
A large part of the inelastic onset  seems ``missing'' in the energy transport. The first question is whether this prediction holds also for \emph{inelastic excitations of different types}, i.e., other than the spin-splitting we studied.
\item\label{diamond}
\emph{The suppression diamond at high bias} inside the Coulomb blockade regime [labeled (vi) in the main article].\\
At the lower boundaries of this regime we found pronounced negative (positive) $\partial I_\text{E}/\partial V$ for positive (negative) $I_\text{E}$. The second question is whether the suppression diamonds \ref{diamond} can still be identified in the qualitatively new Coulomb regimes that can appear due to the different shell-filing possible in the model~\eq{eq:2levelQD}.
\end{enumerate}

\paragraph{Noninteracting multi-level dot:}
In the noninteracting limit $U=U'=0$, which we mention here for completeness (no results shown), our approach also includes the Landauer results for model \eq{eq:2levelQD}.
In fact, as pointed out in \Sec{sec:Landauer} this is the case for \emph{any multi-orbital model}: both concretely in the limit $\Gamma \ll T$, as well as in general for any $\Gamma$ and $T$ when we would renormalize the perturbation theory by appropriately including broadening (not ``by hand''!) while staying in second order.
Such a model does however completely miss the qualitatively new effects due to \icot predicted in the main article which are generated by the intradot Coulomb interaction $U$, see our remark in \Sec{sec:concrete}.

%---------------------------------------------------------
\begin{figure}[t]
\subfloat[]{\includegraphics[height=0.41\textheight]{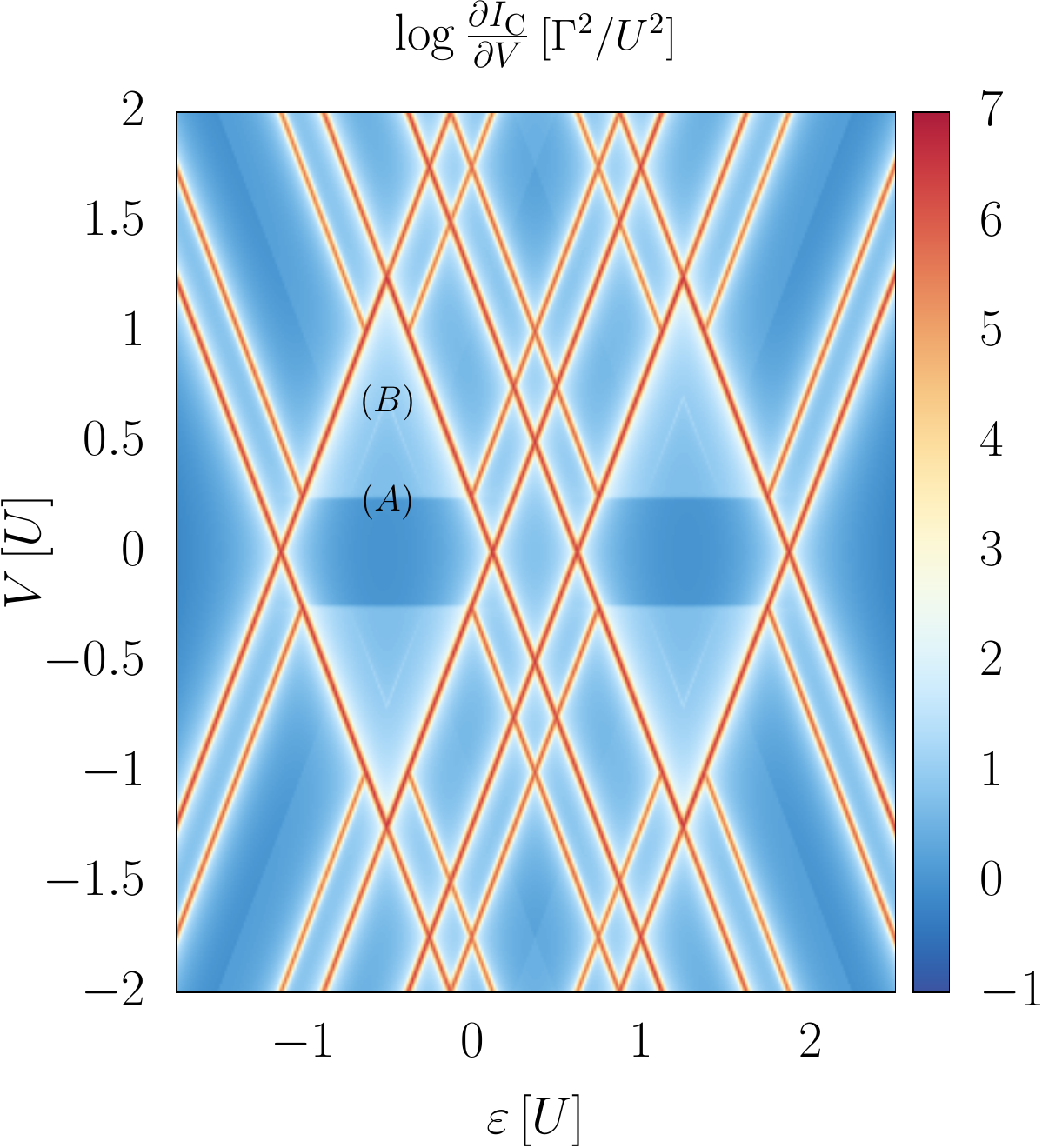}}\hfill
\subfloat[]{\includegraphics[height=0.41\textheight]{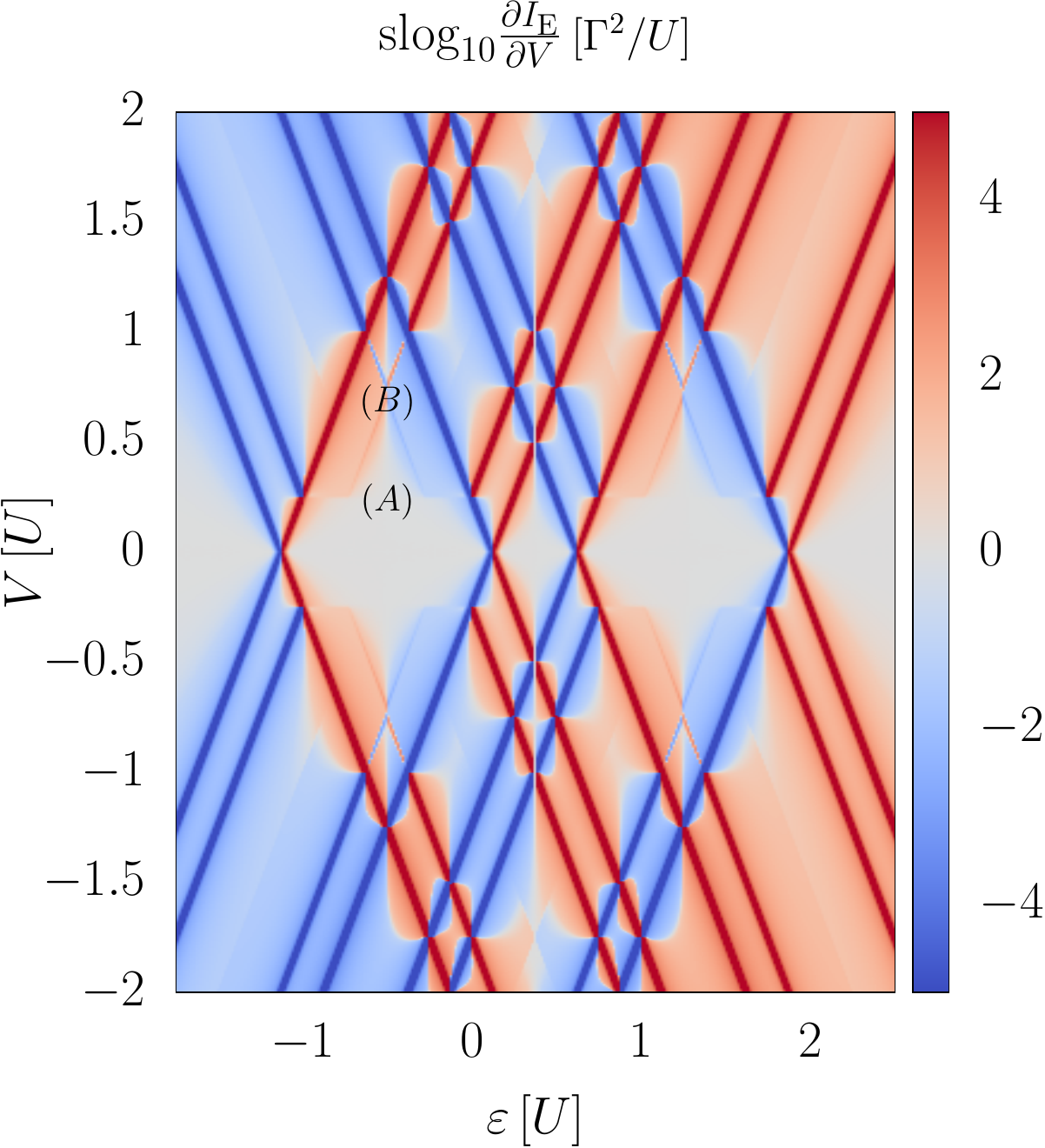}}
\protect\caption{
Transport through a \emph{partially interacting} two-orbital quantum dot with equal intradot interactions 
$U=\frac{1}{3}\cdot\,10^3 \,T$, and vanishing interdot interaction $U'= 0$. 
Of prime interest are the inelastic excitations:
the dots have identical \emph{spin splittings}
$\Delta=  U/4 \approx 83.3 \, T$
but additionally there is a larger \emph{orbital splitting}
$\Delta'= 7U/4 \approx 583.3 \, T$.
All remaining parameters are the same as in the main article:
$\Gamma=\frac{1}{3}\cdot\,10^{-2}\,T$.
Plotted are
(a) the charge conductance $\log_{10} \left( [\partial I_\text{C}/\partial V] / [\Gamma^2/U^2] \right)$
and (b) the energy conductance $\text{slog}_{10}\left( [\partial (I_\text{E}/\partial V)] / [\Gamma^2/U] \right)$
with the signed log, $\text{slog}_{10}(x) := \sgn{x}\log_{10}(a|x|)$ for $a|x|>10$ and linearized $\text{slog}_{10}(x):= ax/10$ for $a|x|<10$ using $a=20$.
\\
Since $\Delta' > 2(U +\Delta)$ the Coulomb blockade regimes do not overlap:
when scanning the ``gate voltage'' $\varepsilon$,
first quantum dot 1 is filled with two electrons and only then dot 2 is filled with two electrons.
The ground states in the five subsequent Coulomb blockade regimes as the level $\varepsilon$ is lowered (going from right to left)
are thus:
$\ket{0                 ; 0}$
$\to$
$\ket{        \downarrow; 0}$ 
$\to$
$\ket{\uparrow\downarrow; 0}$
$\to$
$\ket{\uparrow\downarrow;         \downarrow}$
$\to$
$\ket{\uparrow\downarrow; \uparrow\downarrow}$.
For the $N_1+N_2=1$ and $3$ Coulomb blockade regimes the visible inelastic excitations are due to the excited states
$\ket{        \uparrow; 0}$
and
$\ket{\uparrow\downarrow;         \uparrow}$,
respectively.
\label{fig:multi0}}
\end{figure}

\paragraph{Partially interacting multi-level dot:}
When now each dot is interacting with large intradot charging energy $U \gg \Gamma, T$, but there are is negligible interdot interaction, $U'=0$,
then the two-orbital model reduces to a sum  $H_\text{d} = \sum_{i=1}^2 H_{\text{d},i}$ of commuting Anderson level models, $[H_{\text{d},1},H_{\text{d},2}]=0$.
In principle, this does not yet imply that the currents add up: since each quantum dot still is interacting, the two dots may effectively start to interact \emph{via} the tunneling to the common electrode. However, this is negligible except for special resonant level configurations ($\Delta' \lesssim \Gamma$),  see \Sec{sec:nongeneric}.
In the generic nonresonant case, the current will thus be well approximated by a sum of the currents through the two quantum dots, each strongly interacting, each current as calculated in the main article but with only an orbital offset $\Delta'$.
This is confirmed by explicitly calculated results shown in \Fig{fig:multi0}.
In this case, all the predicted features \ref{triangle} and \ref{diamond} are \emph{replicated precisely once for each orbital}.
%---------------------------------------------------------
\paragraph{Fully interacting multi-level dots:}

\begin{figure}[t]
\subfloat[]{\includegraphics[height=0.41\textheight]{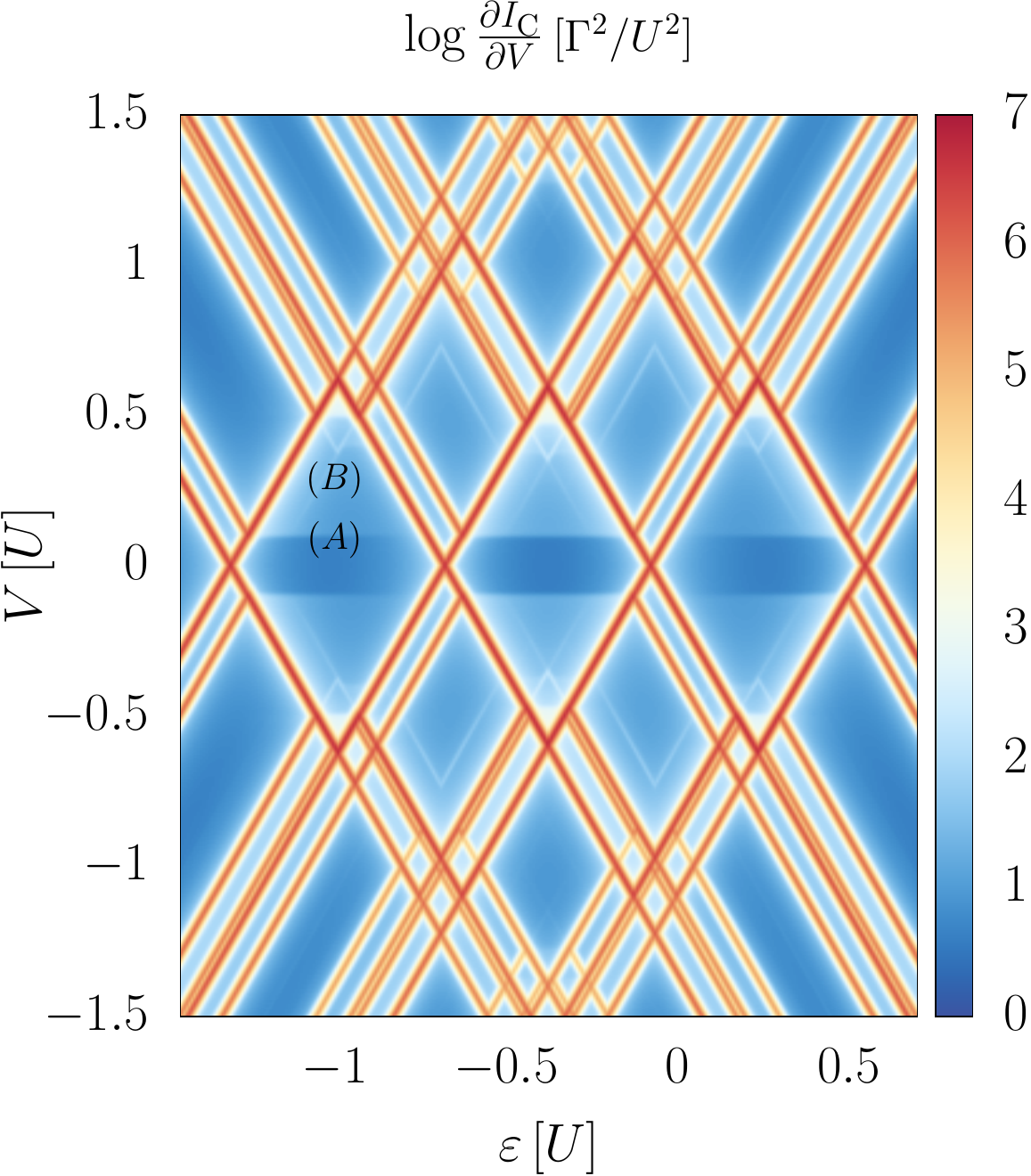}}\hfill
\subfloat[]{\includegraphics[height=0.41\textheight]{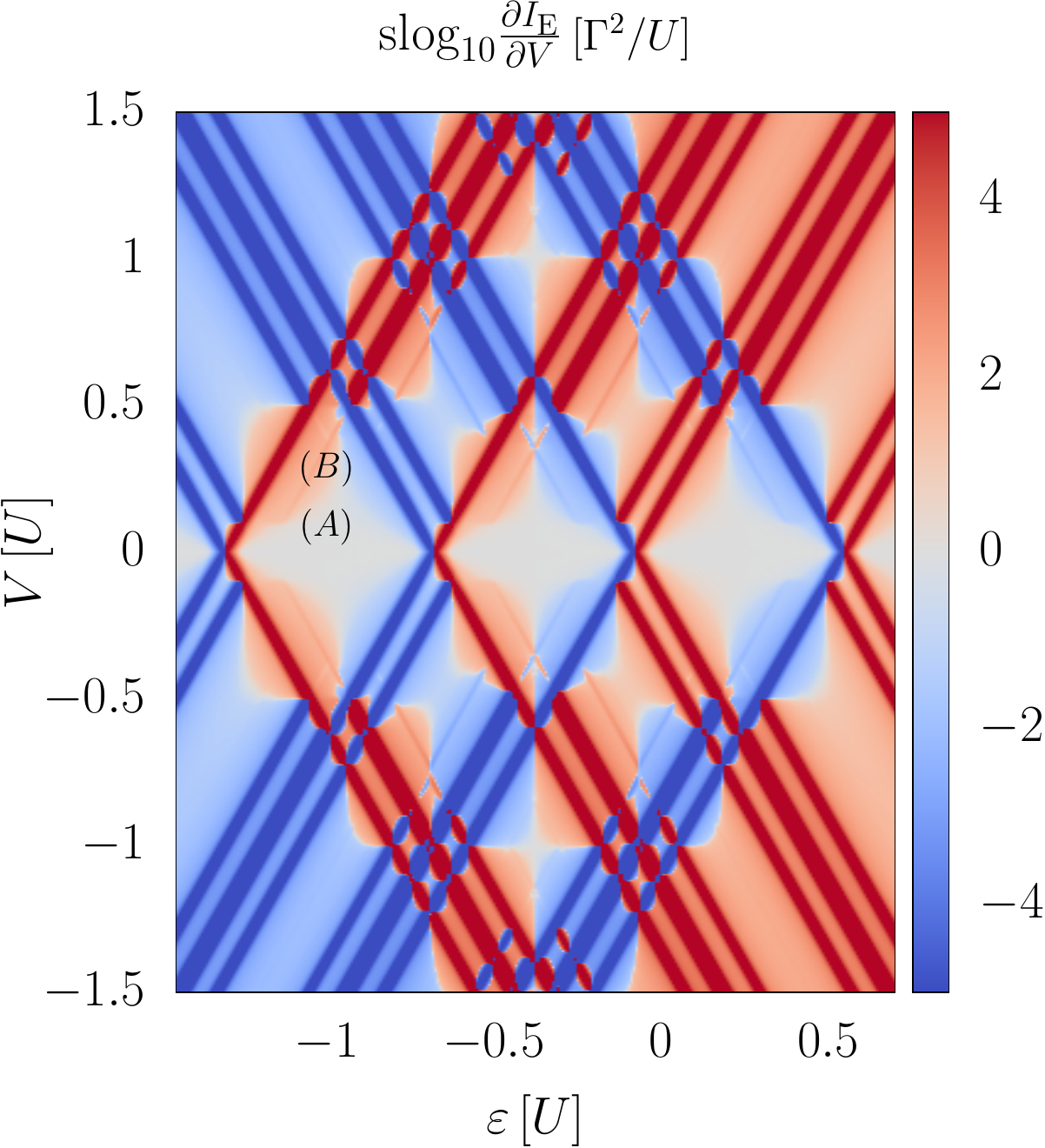}}
\protect\caption{
Transport through a \emph{fully interacting} two-orbital quantum dot,
including both intradot interaction $U = \frac{1}{3}\cdot\,10^3 \,T$, as well as nonzero but small interdot interaction $U'= U/8 \approx 41.7\, T$.
The inelastic excitations due to \emph{spin splittings} $\Delta= U/10 \approx 33.3 \, T$
are still much smaller than \emph{orbital splitting} $\Delta'=  U/2 \approx 166.7 \, T$.
All other parameters and units are the same as in \Fig{fig:multi0}.
\\
Due to the combined effect of large $\Delta'$ and interdot charging $U'$ the Coulomb blockade regimes do not overlap.
However, when scanning the ``gate voltage'' $\varepsilon$
the quantum dots are not filled one after the other as was the case in \Fig{fig:multi0}.
Instead, the ground states in the five subsequent Coulomb blockade regimes as the level $\varepsilon$ is lowered (going from right to left)
are:
$\ket{0                 ; 0}$
$\to$
$\ket{        \downarrow; 0}$ 
$\to$
$\ket{\downarrow; \downarrow}$
$\to$
$\ket{\uparrow\downarrow;         \downarrow}$
$\to$
$\ket{\uparrow\downarrow; \uparrow\downarrow}$.
Now in the central $N_1+N_2=2$ Coulomb blockade regime there is an additional inelastic excitation due to the 2-fold degenerate excited states
$\ket{\uparrow; \downarrow}$, $\ket{\downarrow; \uparrow}$.
Orbital excitations are much higher in energy and can only just be seen in (a) at the top of the middle three Coulomb blockade regimes.
\label{fig:multi1}}
\end{figure}

Now we switch on the \emph{interdot} interaction $U'$. The first example of this kind is shown in \Fig{fig:multi1}. One clearly sees a \emph{single inelastic excitation} relative to the ground state \emph{for each charge state} in three subsequent Coulomb blockade regimes due to the spin splitting $\Delta$ (assumed equal on both dots).
In the central Coulomb blockade regime the ground state is a spin triplet (in our simple model we did not include exchange interaction).
For the charge transport this leads to the standard spectrum, very similar to that calculated~\cite{supKoch06,supLeijnse08a,supStevanato12} and observed experimentally in a variety of systems, qualitatively repeating itself. The COSET features are weakly visible and one notices in the Coulomb blockade regimes adjacent to the central one an asymmetry in the intensities due to the breaking of exact electron-hole symmetry.
Features in the energy conductance appear exactly at the corresponding locations as expected on the basis of the Anderson model calculations of the main article:
\ref{triangle} the inelastic tunneling threshold is missing in the energy conductance, in particular in the central diamond, but also in the adjacent ones;
\ref{diamond} the suppression diamond is most clearly seen in the central Coulomb blockade regimes.
The latter feature also appears in the adjacent Coulomb blockade regimes but only with sharp energy conductance \emph{dips} at its lower boundaries, not negative resonances.
What is more, the inelastic contributions to the energy current are even pronounced \emph{outside} the Coulomb blockade regimes, in particular in the SET regimes between the Coulomb blockade regimes.
Thus the predicted features even tend to \emph{proliferate} in the energy transport spectrum.

\begin{figure}[t]
\subfloat[]{\includegraphics[height=0.41\textheight]{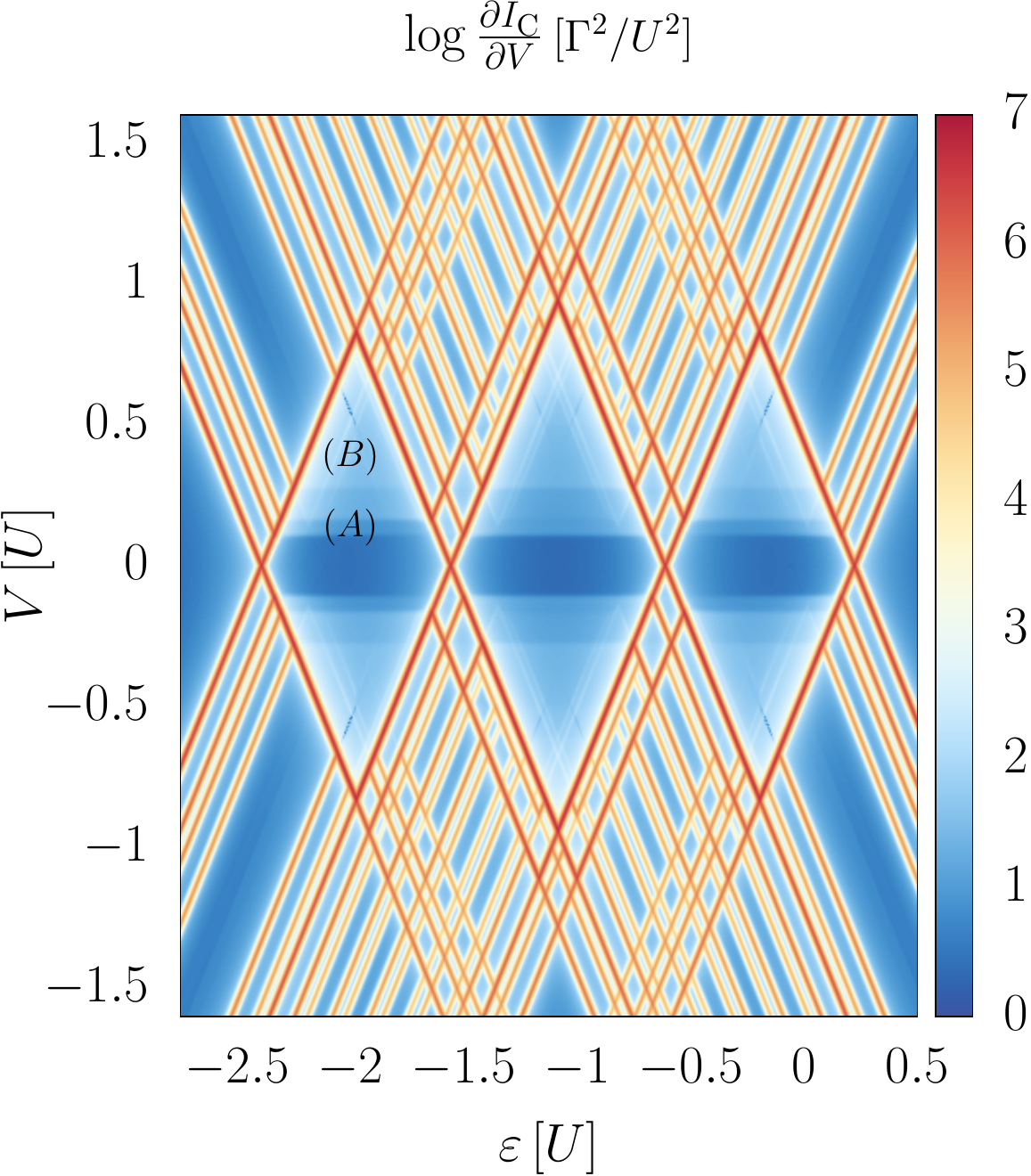}}\hfill
\subfloat[]{\includegraphics[height=0.41\textheight]{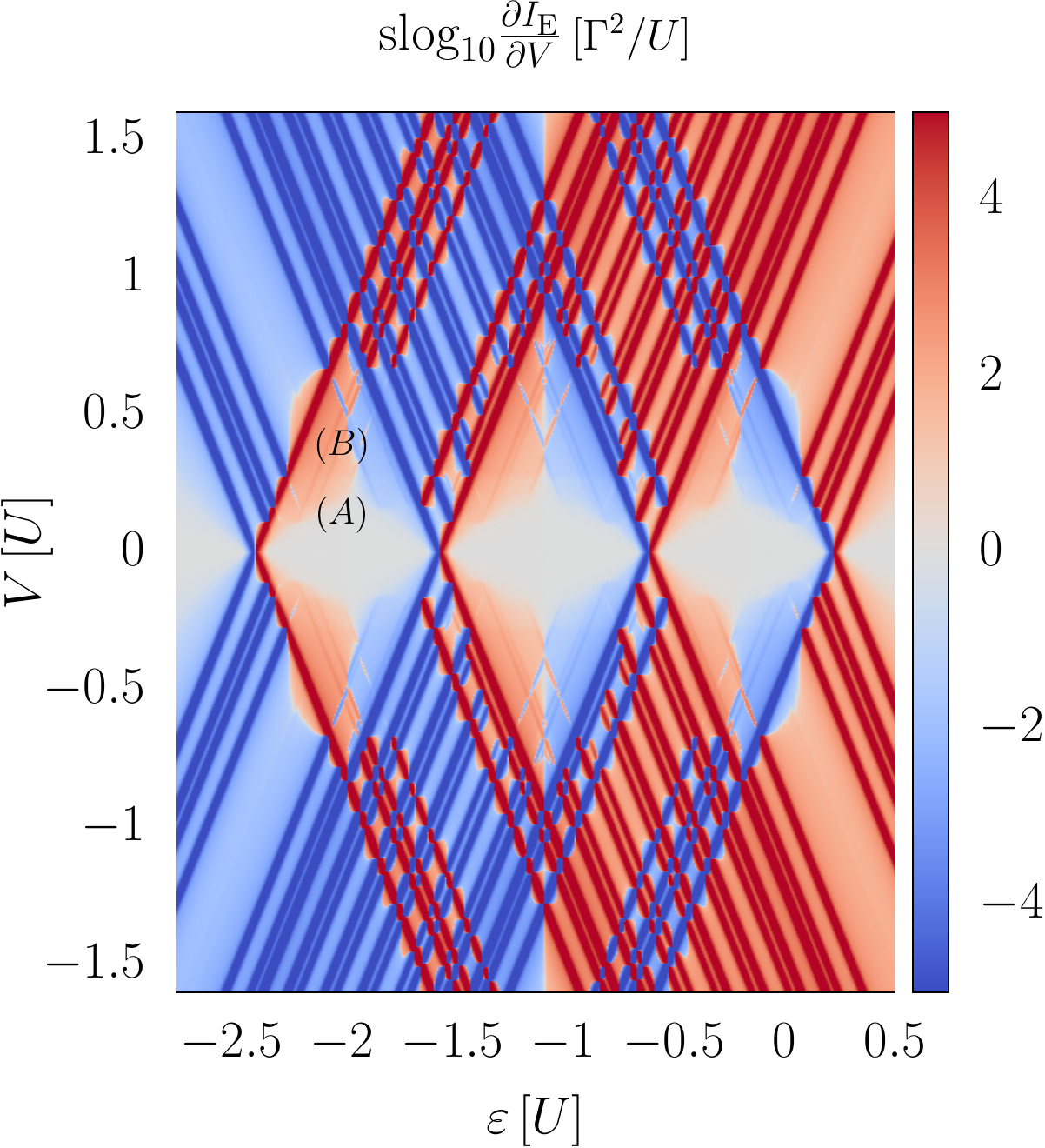}}
\protect\caption{
Transport through a fully interacting two-orbital quantum dot
with intradot interaction $U = \frac{1}{3}\cdot\,10^3 \,T$ and now a relatively strong interdot interaction $U'= 2U/3 \approx 222.2\, T$ as compared to \Fig{fig:multi1}.
Also, the \emph{orbital splitting} $\Delta'=  U/6 \approx 55.6 \, T$ is now much smaller, being
comparable to the \emph{spin splitting} $\Delta= U/9 \approx 37 \, T$.
All other parameters and units are the same as in \Fig{fig:multi0}.
The electron filling of the ground states is the same as in \Fig{fig:multi1}, but the excitations are now all visible in the central three Coulomb blockade regimes.
\label{fig:multi3}}
\end{figure}

In the second example of a fully interacting two-orbital model, shown in \Fig{fig:multi3},  we now see \emph{several inelastic excitations in one charge state}, due to both \emph{spin} and \emph{orbital} excitations.
It can be clearly seen for each of these that the inelastic threshold is missing in the energy conductance, i.e., the predicted suppression triangle \ref{triangle} is \emph{also replicated within one charge state}.
Moreover, the suppression diamond \ref{diamond} is also replicated: in the central Coulomb blockade regime two such diamonds are clearly visible with pronounced negative $\partial I_\text{E}/\partial V$ at their boundaries.
In the left and right Coulomb blockade regimes single suppression diamonds \ref{diamond} with clear negative $\partial I_\text{E}/\partial V$ are seen.
However, in addition negative $\partial I_\text{E}/\partial V$ is seen even at several COSET features at lower bias, i.e., this feature can also \emph{proliferate}. Finally, this figure also illustrates the statement in the conclusion of the main article that the regime where approximations that neglect the charge fluctuations involved in COSET are valid rapidly shrinks when more inelastic excitations are present: a large part of the Coulomb blockade regime shows the COSET energy transport of the type predicted in the article.

\subsubsection{Expected deviations from generic energy current effects\label{sec:nongeneric}}

The above examples confirm that the effects predicted in the main article are quite generic for \emph{multi-orbital} models in the sense that they are replicated for each orbital level as the gate voltage is scanned. In fact, the observation that the energy transport effects tends to proliferate while the charge transport remains featureless further strengthens a central point of the main paper: the energy conductance is far more sensitive and thus a promising new experimental tool.
Although we studied only two orbitals, adding further orbitals does not bring anything qualitatively new. Only the intensities of inelastic excitations may vary, which are governed by the many-body quantum state of the dot in combination with the orbital specific tunneling matrix elements. 
Finally, within our general approach it also is straightforward to identify situations in which we can expect interesting deviations from the generic behavior:
\begin{itemize}
\item
As mentioned in the discussion of the main article and in \Sec{sec:Scaling-analysis}, at low temperature \icot processes are significantly enhanced by renormalization effects and are generally expected to make the inelastic energy transport effects more pronounced.
Charge fluctuations already enhance the low-temperature \icot as shown in Ref.~\onlinecite{supSaptsov14a} and most importantly spin-fluctuations will further renormalize the {\icot}.
Extending the considerations in \Sec{sec:Scaling-analysis} using naive poor-man scaling arguments
the conductance at $V=\Delta$ becomes logarithmically enhanced to
$\partial I_\text{C} /\partial V \sim \Gamma^2/U^2 \left( \ln(T/T_K) \right)^{-1}$
as $T$ approaches the Kondo scale $T_K$ from above ($T\gg T_K$) and we expect that at this voltage the energy current also experiences an enhancement.
It is an interesting question how the cancellations of energy current contributions described in the main article develop under this renormalization.
\item
When a quantum dot has degenerate levels that are not protected by symmetries of the total system (i.e., dot plus electrodes)
the master equation \eq{eq:master-equation} must be extended to a \emph{quantum master equation} which includes the coupling of occupations to coherences of the density matrix in the energy basis. This is, however, not a standard quantum master equation since it requires the coherences to be calculated to $\mathcal{O}(\Gamma)+\mathcal{O}(\Gamma^2)$. Such equations can be derived systematically within our general approach, see \Cite{supHell14b}. To  $\mathcal{O}(\Gamma)$ thermoelectric effects involving quantum dots with spin degeneracies coupled to spin-polarized electrodes (spin-valve junctions) have been studied in the \set regime~\cite{supMuralidharan13}.
Degeneracy effect are important also for orbital degeneracies and have been studied without energy transport, e.g., in double quantum dots~\cite{supWunsch05}, single-atom~\cite{supDonarini12b} or single-molecule junctions~\cite{supDarau09}.

\end{itemize}
It is an interesting open question how these deviations will appear on the background of the generic effects identified in the main article.


\begin{thebibliography}{1}
\bibitem{supfootnote}
One can show that due to the restriction to probabilities the contribution $L$ to $L_{\text{eff}}$ in \Eq{eq:perturbative-series} drops out.
\end{thebibliography}
\end{document}